\documentclass[showpacs,preprintnumbers,amsmath,amssymb,twocolumn,superscriptaddress,prb,floatfix]{revtex4}
\usepackage{times}
\usepackage{graphicx}
\usepackage{amsfonts}
\usepackage{amsmath, amsthm, amssymb}
\usepackage{dsfont}

\newcommand{\A}{{\cal{A}}_{\mathbf{k}}}         						
\newcommand{\Imag}{{\mathrm{Im}}}   												
\def\i{\mathrm{i}}  
\newcommand{\im}{\mathrm{i}}        												
\newcommand{\ve}[1]{\mathbf{#1}}
\DeclareMathOperator{\diag}{diag} 													
\newcommand{\x}{\lambda}  																	
\newcommand{\y}{\rho}     																	
\newcommand{\T}{\mathrm{T}}   															
\newcommand{\vk}{\ve{k}} 																		
\newcommand{\vp}{\ve{p}} 																		
\newcommand{\vq}{\ve{q}} 																		
\newcommand{\ca}[2][]{c_{#2}^{\vphantom{\dagger}#1}} 				
\newcommand{\cc}[2][]{c_{#2}^{{\dagger}#1}}          				
\newcommand{\da}[2][]{d_{#2}^{\vphantom{\dagger}#1}} 				
\newcommand{\dc}[2][]{d_{#2}^{{\dagger}#1}}          				
\newcommand{\ga}[2][]{\gamma_{#2}^{\vphantom{\dagger}#1}} 	
\newcommand{\gc}[2][]{\gamma_{#2}^{{\dagger}#1}}          	
\newcommand{\Tkp}[1]{T_{\vk\vp#1}}  												

\newcommand{\e}[1]{\mathrm{e}^{#1}}
\newcommand{\dif}{\mathrm{d}} 															
\newcommand{\mean}[1]{\langle#1\rangle}
\newcommand{\abs}[1]{|#1|}

\newcommand{\pauli}[1][\alpha\beta]{\hat{\boldsymbol{\sigma}}_{#1}^{\vphantom{\dagger}}}
\newcommand{\pauliDAG}[1][\beta\alpha]{(\hat{\boldsymbol{\sigma}}_{#1})^{\ast}}
\newcommand{\pauliN}{\hat{\boldsymbol{\sigma}}^{\vphantom{\dagger}}}
\newcommand{\wdf}[2][1/2]{{\hat{\cal{D}}}_{#2}^{(#1)}(\vartheta)} 				
\newcommand{\TT}{\tilde{\mathrm{T}}}
\newcommand{\wm}{\omega_m} 
\newcommand{\wnu}{\widetilde{\omega}_\nu}
\newcommand{\I}{\mathbf{I}^\text{S}}   
\newcommand{\bPhi}{\mathbf{\Phi}}  
\newcommand{\bk}{B_{\vk,-}}
\newcommand{\bp}{B_{\vp,-}}

\newcommand{\ie}{\textit{i.e. }}		
\newcommand{\eg}{\textit{e.g. }}	
\newcommand{\etal}{\emph{et al. }}

\begin{document}
\title[Tunneling currents in ferromagnetic systems with multiple broken symmetries]{Tunneling currents in ferromagnetic systems with multiple broken symmetries}
\author{J. Linder}
\affiliation{Department of Physics, Norwegian University of
Science and Technology, N-7491 Trondheim, Norway}
\author{M. S. Gr{\o}nsleth}
\affiliation{Department of Physics, Norwegian University of
Science and Technology, N-7491 Trondheim, Norway}
\author{A. Sudb{\o}}
\affiliation{Department of Physics, Norwegian University of
Science and Technology, N-7491 Trondheim, Norway}
\affiliation{Centre for Advanced Study, Norwegian Academy of Science and Letters, Drammensveien 78, N-0271 Oslo, Norway.  }
\date{Received \today}
\begin{abstract}
  A system exhibiting multiple simultaneously broken symmetries offers
  the opportunity to influence physical phenomena such as tunneling
  currents by means of external control parameters. In this paper, we consider
  the broken $SU(2)$ (internal spin) symmetry of ferromagnetic systems coexisting with \textit{i)} the 
  broken $U(1)$ symmetry of superconductors and \textit{ii)} the broken spatial inversion
  symmetry induced by a Rashba term in a spin-orbit coupling Hamiltonian. In
  order to study the effect of these broken symmetries, we consider
  tunneling currents that arise in two different systems;
  tunneling junctions consisting of non-unitary spin-triplet
  ferromagnetic superconductors and junctions consisting of
  ferromagnets with spin-orbit coupling. In the former case, we
  consider different pairing symmetries in a model where
  ferromagnetism and superconductivity coexist uniformly. 
  An interplay between the relative magnetization orientation 
  on each side of the junction and the superconducting phase
  difference is found, similarly to that found in earlier
  studies on spin-singlet superconductivity coexisting with 
  spiral magnetism. This interplay 
  gives rise to persistent spin- and charge-currents in 
  the absence of an electrostatic voltage that can be 
  controlled by adjusting the relative magnetization 
  orientation on each side of the junction. In the second 
  system, we study transport of spin in a system consisting 
  of two ferromagnets with spin-orbit coupling
  separated by an insulating tunneling junction. A persistent
  spin-current across the junction is found, which can be 
  controlled in a well-defined manner by external magnetic 
  and electric fields. The behavior of this spin-current 
  for important geometries and limits is studied.
\end{abstract}

\maketitle

\section{Introduction}
Due to the increasing interest in the field of spintronics in recent
years \cite{zutic2004}, the idea of utilizing the spin degree of
freedom in electronic devices has triggered an extensive response in
many scientific communities. The spin-Hall effect is arguably the
research area which has received most focus in this context, with
substantial effort being put into theoretical considerations
\cite{dyakonov1971} as well as experimental observations
\cite{wunderlich2005}. In spintronics, a main goal is to make use of the
spin degree of freedom rather than electrical charge, investigations
of mechanisms that offer ways of controlling spin-currents are of
great interest. The study of systems with multiple broken
symmetries is highly relevant in this context, since such systems
promise rich physics with the opportunity to learn if the tunneling
currents can be influenced by means of external control parameters
such as electric and/or magnetic fields. Here, we will focus on two
specific systems: ferromagnetism coexisting with
superconductivity, which we shall refer to as ferromagnetic
superconductors (FMSC), and systems where ferromagnetism and
spin-orbit coupling are present (FMSO). In terms of broken symmetries, we will 
then study the broken $SU(2)$ (internal spin) symmetry of ferromagnetic systems coexisting with 
the broken $U(1)$ symmetry of superconductors and also consider ferromagnets with broken 
inversion (spatial)
symmetry induced by a Rashba term in a spin-orbit coupling Hamiltonian.

\par
The coexistence of ferromagnetism (FM) and superconductivity (SC) has
a short history in experimental physics
\cite{saxena2000,aoki2001,tallon1999}, although a
theoretical proposition of this phenomenon was offered as early as
1957 by Ginzburg \cite{ginzburg1957}. Spin-singlet superconductivity
originating with BCS theory seems to be ruled out as a plausible
pairing mechanism for a ferromagnetic superconductor \cite{shen2003},
at least with regard to uniform coexistence of the FM and SC order
parameters $\zeta$ and $\Delta$, respectively. It could be achieved
for a superconductor taking up a so-called
Fulde-Ferrell-Larkin-Ovchinnikov (FFLO) state \cite{fflo}.
However, it seems likely that the coexistence of FM and SC call for
\cite{walker2002,machida2001} $p$-wave spin-triplet Cooper pairs which
have a non-zero magnetic moment. This type of pairing has been
observed in superfluid $^3$He, and is perfectly compatible with FM
order.  Spin-triplet superconductivity has moreover been
experimentally verified \cite{ishida1998, nelson2004} in Sr$_2$RuO$_4$, and the study of such a pairing in a FMSC could unveil interesting effects
with respect to quantum transport. The concept of simultaneously
broken $U(1)$ and $SU(2)$ symmetries are of great interest from a
fundamental physics point of view, and could be suggestive to a range of novel applications.  This topic has been the subject 
of theoretical research in {\it e.g.}
Refs.~\onlinecite{brataas2004,tanaka2004,koyama1984}.

\par
In this paper, we follow up
Ref.~\onlinecite{gronsleth} with a more comprehensive
study of the tunneling currents between two $p$-wave FMSC separated by
an insulating junction; RuSr$_2$GdCu$_2$O$_8$, UGe$_2$, and
URhGe have been proposed as candidates for such unconventional
superconductors \cite{tallon1999, saxena2000, aoki2001}. 
In our model, we assume uniform coexistence of the FM and
SC order parameters and that superconductivity arises from 
the same electrons that are responsible for the magnetism. 
As argued in Ref.~\onlinecite{saxena2000}, this can be 
understood most naturally as a spin-triplet rather than spin-singlet pairing phenomenon. Furthermore, it seems that 
SC in the metallic compounds mentioned above always coexists 
with the FM order and is enhanced by it \cite{shopova2005}; 
the experiments conducted on the compounds
UGe$_2$ and URhGe do not give any evidence for the
existence of a standard normal-to-superconducting phase 
transition in a zero external magnetic field, but instead 
indicate a phase corresponding to a mixed state of FM and 
SC. We provide detailed calculations for single-particle 
and Josephson (two-particle) tunneling between two 
non-unitary equal-spin pairing (ESP) FMSC. We examine both 
the charge- and spin-sector in detail within linear
response theory using the Kubo formula. We find that the supercurrent of spin and charge may be controlled by adjusting the misorientation of the exchange fields on both sides of the junction. Such an effect was first discovered by Kulic and Kulic \cite{kulic2005}, who derived an expression for the Josephson current over a junction separating two BCS superconductors with spiral magnetic order. It was found that the supercurrent could be controlled by adjusting the relative orientation of the exchange field on both sides of the junction, a finding that quite remarkably suggested a way of tuning a supercurrent in a well-defined manner from \eg a 0- to $\pi$-junction. Later investigations made by Eremin, Nogueira, and Tarento \cite{eremin2006} considered a similar system as 
Kulic and Kulic \cite{kulic2005}, namely two Fulde-Ferrel-Larkin-Ovchinnikov (FFLO) superconductors \cite{fflo} coexisting with helimagnetic order. Recently, the same opportunity was found to exist in a FMSC/I/FMSC junction as shown by Gr{\o}nsleth \etal \cite{gronsleth}. 
\par
In the case of a system where both ferromagnetism and
spin-orbit coupling are present, it is clear that these are physical
properties of a system that crucially influence the behavior of spins
present in that system. For instance, the presence of spin-orbit
coupling is highly important when considering ferromagnetic
semiconductors \cite{dietl2002, matsukura2002}. Such materials have been
proposed as devices for obtaining controllable spin injection and
manipulating single electron spins by means of external electrical
fields, making them a central topic of semiconductor spintronics
\cite{engel2006}. In ferromagnetic metals, spin-orbit coupling is
ordinarily significantly smaller than for semiconductors due to the
bandstructure. However, the presence of a spin-orbit coupling in
ferromagnets could lead to new effects in terms of quantum transport.

\par
Studies of tunneling between ferromagnets have uncovered interesting
physical effects \cite{nogueira2004b, lee2003, slonczewski1989}.
Nogueira \etal~predicted \cite{nogueira2004b} that a dissipationless
spin-current should be established across the junction of two
Heisenberg ferromagnets, and that the spin-current was maximal in the
special case of tunneling between planar ferromagnets. Also, there has
been investigations of what kind of impact spin-orbit coupling
constitutes on tunneling currents in various contexts, \eg for
noncentrosymmetric superconductors \cite{yokoyama2005}, and
two-dimensional electron gases coupled to ferromagnets
\cite{wang2005}. Broken time reversal- and inversion-symmetry are
interesting properties of a system with regard to quantum transport of
spin and charge, and the exploitment of such asymmetries has given
rise to several devices in recent years. For instance, the broken $SU(2)$
symmetry exhibited by ferromagnets has a broad range of possible
applications. This has led to spin current induced magnetization
switching \cite{kiselev2003}, and suggestions have been made for more
exotic devices such as spin-torque transistors \cite{bauer2003} and
spin-batteries \cite{brataas2002}. It has also led to investigations
into such phenomena as spin-Hall effect in paramagnetic metals
\cite{hirsch1999}, spin-pumping from ferromagnets into metals,
enhanced damping of spins when spins are pumped from one ferromagnet
to another through a metallic sample \cite{tserkovnyak2002}, and the
mentioned spin Josephson effects in ferromagnet/ferromagnet tunneling
junctions \cite{nogueira2004b}.

\par
Here, we study the spin-current that arises over a tunneling
junction separating two ferromagnetic metals with substantial
spin-orbit coupling. It is found that the total current consists of three
terms; one due to a twist in magnetization across the junction (in
agreement with the result of Ref.~\onlinecite{nogueira2004b}), one term
originating from the spin-orbit interactions in the system, and
finally an interesting mixed term that stems from an interplay between
the ferromagnetism and spin-orbit coupling. After deriving the
expression for the spin-current between Heisenberg ferromagnets with
substantial spin-orbit coupling, we consider important tunneling
geometries and physical limits of our generally valid results.
Finally, we make suggestions concerning the detection of the predicted
spin-current. Our results indicate how spin transport between systems
exhibiting both magnetism and spin-orbit coupling can be controlled by
external fields, and should therefore be of considerable interest in
terms of spintronics.

\par
This paper is organized as follows. In Sec. \ref{sec:fmsc}, we
consider transport between spin-triplet ferromagnetic superconductors,
while a study of transport between ferromagnets with spin-orbit
coupling is given in Sec. \ref{sec:fmsoc}. A discussion of our results
is provided in Sec. \ref{sec:discuss}, with emphasis on how the novel
effects predicted in this paper could be tested in an experimental
setup.  Finally, we give concluding remarks in Sec. \ref{sec:summary}.

\section{Ferromagnetic superconductors}\label{sec:fmsc}

\subsection{Coexistence of ferromagnetism and superconductivity}\label{sec:uniform}
An important issue to address concerning FMSC is whether the SC and FM
order parameters coexist uniformly or if they are phase-separated. One
possibility \cite{tewari2004} is that a spontaneously formed vortex
lattice due to the internal magnetization $\mathbf{m}$ is realized in
a spin-triplet FMSC, while there also have been studies of Meissner
(uniform) SC phases in spin-triplet FMSC \cite{shopova2005}. As argued
in Ref.~\onlinecite{mineev2005}, a key variable with respect to
whether a vortice lattice appears or not is the strength of the
internal magnetization $\mathbf{m}$. Ref.~\onlinecite{mineev1999}
suggested that vortices arise if $4\pi\mathbf{m}>\mathbf{H}_{c1}$, where
$\mathbf{H}_{c1}$ is the lower critical field. When considering a weak
FM state coexisting with SC, a scenario which seems to be the case for
URhGe, the domain structure in the absence of an external
field is thus vortex-free. Current experimental data concerning
URhGe are not strong enough to unambiguously settle this
question, while evidence for uniform coexistence of FM and SC has been
indicated \cite{kotegawa2005} in UGe$_2$.  Furthermore, a bulk
Meissner state in the FMSC RuSr$_2$GdCu$_2$O$_8$ has been reported in
Ref.~\onlinecite{bernhard2000}, hence suggesting the existence of
uniform FM and SC as a bulk effect. In our study, we shall
consequently take the order parameters as coexisting homogeneously and
use their bulk values, as justified by the argumentation above.
However, we emphasize that one in general should take into account the
possible suppression of the SC order parameter in the vicinity of the
tunneling interface due to the formation of midgap surface states
\cite{hu1994} which occur for certain orientations of the SC gap. The
pair-breaking effect of these states in unconventional superconductors
has been studied in e.g. \cite{ambegaokar1974, buchholtz1981,
  tanuma2001}, and we discuss this in more detail in Sec.
\ref{sec:discuss}. A sizeable formation of such states would suppress
the Josephson current, although it is nonvanishing in the general
case. Also, we use bulk uniform magnetic order parameters, as in
\cite{nogueira2004b}. The latter is justified on the
grounds that a ferromagnet with a planar order parameter is
mathematically isomorphic to an $s$-wave superconductor, where the use
of bulk values for the order parameter right up to the interface is a
good approximation due to the lack of midgap surface states.

\par
\begin{figure}[!ht]
\centering
\resizebox{0.42\textwidth}{!}{
\includegraphics{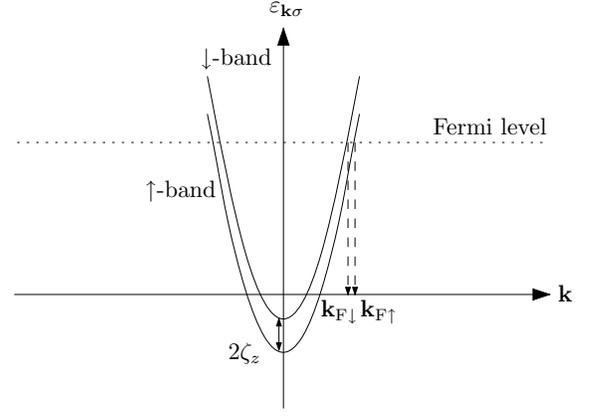}}
\caption{Band-splitting for $\uparrow,\downarrow$ electrons in the
  presence of a magnetization in
  $\hat{\mathbf{z}}$-direction. Inter-band pairing gives rise to a net
  Cooper pair momentum in the presence of a band-splitting, thus
  suppressing the $\Delta_{\vk\uparrow\downarrow}$ order parameter.}
\label{fig:split}
\end{figure}

It is generally believed that the same electrons that are responsible
for itinerant FM also participate in the formation of Cooper pairs below the SC
critical temperature \cite{aoki2001}. As a consequence,
uniform coexistence of spin-singlet SC and FM can be discarded since
$s$-wave Cooper pairs carry a total spin of zero, although spatially
modulated order parameters could allow for magnetic $s$-wave
superconductors \cite{kulic2005,eremin2006}. However, spin-triplet
Cooper pairs are in principle perfectly compatible with FM order since
they can carry a net magnetic moment. To see this, consider the
$\mathbf{d}_\vk$-vector formalism\cite{datta1990} which is convenient
when dealing with spin-triplet superconductors, regardless of whether
they are magnetic or not. For a complete and rigorous treatment of the
$\mathbf{d}_\vk$-vector order parameter, see \eg Ref.
~\onlinecite{leggett1975}. The spin dependence of triplet pairing can
be represented by a 2$\times$2 matrix
\begin{align}
  \hat{\Delta}_\vk &= \begin{pmatrix}
    \Delta_{\vk\uparrow\uparrow} & \Delta_{\vk\uparrow\downarrow}\\
    \Delta_{\vk\downarrow\uparrow} & \Delta_{\vk\downarrow\downarrow}
  \end{pmatrix} 
  \\ \nonumber
  &= \begin{pmatrix}
    -d_x(\vk) +\i d_y(\vk) & d_z(\vk)\\
    d_z(\vk) & d_x(\vk) +\i d_y(\vk)
  \end{pmatrix}
  = \i\mathbf{d}_\vk\cdot\pauliN\hat{\sigma}_y,
\end{align}
where $\Delta_{\vk\alpha\beta}$ represent the SC gap parameters for
different triplet pairings, $\pauliN = (\hat{\sigma}_x,\hat{\sigma}_y,\hat{\sigma}_z)$
where $\hat{\sigma}_i$ are the Pauli matrices, and
$\mathbf{d}_\vk = (d_x(\vk),d_y(\vk),d_z(\vk))$ is given by
\begin{equation}
  \mathbf{d}_\vk = \Big( 
  \frac{\Delta_{\vk\downarrow\downarrow}-\Delta_{\vk\uparrow\uparrow}}{2}, 
  -\i\frac{(\Delta_{\vk\downarrow\downarrow}+\Delta_{\vk\uparrow\uparrow})}{2}, 
  \Delta_{\vk\uparrow\downarrow}\Big).
\end{equation}
Note that $\mathbf{d}_\vk$ transforms like a vector under spin
rotations and that $\Delta_{\vk\uparrow\downarrow} =
\Delta_{\vk\downarrow\uparrow}$ for triplet pairing since it is of no
significance \textit{which} electron in the Cooper pair that has spin
up or down. This is because spin-part of the two-particle wavefunction is symmetric under exchange of particles, as opposed to spin-singlet SC, where the gap changes sign when the spin indices are exchanged. Spin-triplet SC states are classified as unitary if
$\i\mathbf{d}_\vk\times\mathbf{d}_\vk^*=0$ and non-unitary if the
equality sign does not hold. Since the average spin of a
$\mathbf{d}_\vk$-state is given by \cite{leggett1975}
\begin{equation}\label{eq:Cooperpairspin}
  \langle\mathbf{S}_\vk\rangle = \i\mathbf{d}_\vk\times\mathbf{d}^*_\vk,
\end{equation}
it is clear that we must have a non-unitary $\mathbf{d}_\vk$ in a model
where FM and SC coexist uniformly. Indeed, there is strong reason to
believe that the correct pairing symmetries in the discovered FMSC
constitute non-unitary states \cite{hardy2005,samokhin2002,
  machida2001}. As a consequence, one can rule out for instance a
state where only $\Delta_{\vk\uparrow\downarrow}\neq0$ since it would
imply $\langle\mathbf{S}_\vk\rangle=0$ according to Eq.
(\ref{eq:Cooperpairspin}). In the most general case where all SC gaps
are included, $\Delta_{\uparrow\downarrow}$ would be suppressed in the
presence of a Zeeman-splitting between the $\uparrow, \downarrow$
conduction bands \cite{aoki2001}; see Fig. \ref{fig:split}.

\begin{figure}[!ht]
  \centering
  \resizebox{0.42\textwidth}{!}{
    \includegraphics{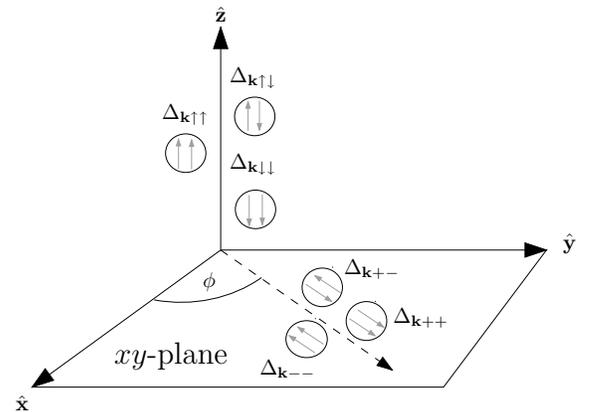}}
  \caption{Change of spin-basis for the superconducting gaps. The new
    quantization axis is represented by the dotted arrow.}
  \label{fig:gap}
\end{figure}

However, such a splitting between energy-bands need not be present and
one could in theory then consider a $\mathbf{d}$-vector where
\begin{equation}\label{eq:condxy}
 |\Delta_{\vk\uparrow\uparrow}|=|\Delta_{\vk\downarrow\downarrow}|\neq0,\;\Delta_{\vk\uparrow\downarrow}\neq 0
\end{equation}
such that $\langle\mathbf{S}_\vk\rangle$ lies in the local $xy$-plane.
This scenario would be equivalent to an A2-phase as is seen when
performing a spin rotation on the gap parameters into a quantization
axis lying in the $xy$-plane. Denoting up- and down-spins with respect
to the new quantization axis by $+$ and $-$, respectively, the
transformation yields
\begin{align}\label{eq:rotmat}
  \begin{pmatrix}
    \Delta_{\vk\uparrow\uparrow}\\
    \Delta_{\vk\uparrow\downarrow}\\
    \Delta_{\vk\downarrow\downarrow}
  \end{pmatrix} =
  \frac{1}{2}
  \begin{pmatrix}
    1 & 2\e{\i\phi} & \e{2\i\phi} \\
    -\e{-\i\phi} & 0 & \e{\i\phi} \\
    \e{-2\i\phi} & -2\e{-\i\phi} & 1 
  \end{pmatrix}
  \begin{pmatrix}
    \widetilde{\Delta}_{\vk++}\\
    \widetilde{\Delta}_{\vk+-}\\
    \widetilde{\Delta}_{\vk--}
  \end{pmatrix}, 
\end{align}
where $\phi$ is the azimuthal angle as shown in Fig. \ref{fig:gap}. 

When introducing the conditions in Eq. (\ref{eq:condxy}), it is
readily seen that $\widetilde{\Delta}_{\vk+-}=0$ while
$|\widetilde{\Delta}_{\vk++}|\neq|\widetilde{\Delta}_{\vk--}|\neq0$,
thus corresponding to an A2-phase. \emph{Consequently, the entire span
  of physically possible pairing symmetries in a FMSC can be reduced
  to the equivalence of an A1- or A2-phase in $^3$He by a change of
  spin-basis.}  The definitions of A-, A1-, and A2-phases in $^3$He
are as follow: an A-phase corresponds to a pairing symmetry such that
$|\Delta_{\vk\uparrow\uparrow}|=|\Delta_{\vk\downarrow\downarrow}|\neq
0$, an A1-phase has only one gap $\Delta_{\vk\sigma\sigma}\neq 0$
while $\Delta_{\vk,-\sigma,-\sigma}=0$, and an A2-phase satisfies
$|\Delta_{\vk\uparrow\uparrow}|\neq|\Delta_{\vk\downarrow\downarrow}|\neq
0$. In this case, $\Delta_{\vk\alpha\beta}$ represents the superfluid
gap for the fermionic $^3$He-atoms, and
$\Delta_{\vk\uparrow\downarrow}=0$ for all $A_i$-phases.

\par
The resulting spin of the Cooper pair is then in general given by

\begin{equation}\label{eq:spinA2}
  \langle\mathbf{S}_\vk\rangle = (1/2)[|\Delta_{\vk\uparrow\uparrow}|^2 - |\Delta_{\vk\downarrow\downarrow}|^2]\hat{\mathbf{z}}.
\end{equation}

\par
\begin{figure}[!ht]
\centering
\resizebox{0.48\textwidth}{!}{
\includegraphics{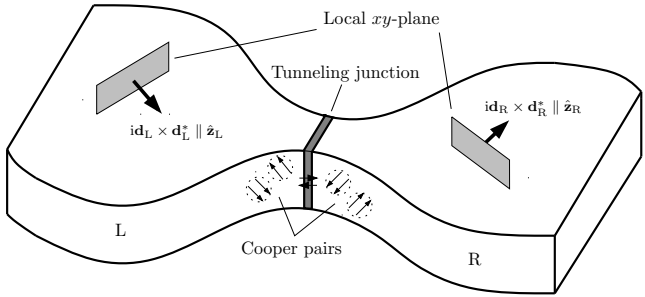}}
\caption{Tunneling between two non-unitary ESP FMSC. The quantization
  axis has been taken along the direction of magnetization on each
  side of the junction.}
\label{fig:junction}
\end{figure}

In the following, we shall accordingly consider tunneling between
non-unitary ESP FMSC in an A1- or A2-phase. Moreover, we consider thin
film FMSC, ensuring that no accumulation of charge at the surface will
take place due to an orbital effect. Our system can be thought to have
arisen by first cooling down a sample below the Curie temperature
$T_\text{M}$ such that FM order is introduced. At further cooling below the
critical temperature $T_\text{c}$, the same electrons that give rise to FM
condense into Cooper pairs with a net magnetic moment parallel to the
original direction of magnetization. Our model is shown in Fig.
\ref{fig:junction}.

\subsection{The Hamiltonian}\label{sec:hamiltonian}

\par The system consists of two FMSC separated by an insulating layer
such that the total Hamiltonian can be written as \cite{cohen1962} $H
= H_\text{L} + H_\text{R} + H_\text{T},$ where L and R represents the
individual FMSC on each side of the tunneling junction, and
$H_\text{T}$ describes tunneling of particles through the insulating
layer separating the two pieces of bulk material. Using mean-field
theory, one finds that the individual FMSC are described by a
Hamiltonian similar to the one used in Ref.~\onlinecite{dahal2005},
\begin{align}
  \label{eq:1}
  H_\text{FMSC} &= H_0 + \sum_{\vk}\psi_{\vk}^\dag
  \hat{\cal{A}}_{\vk}\psi_{\vk},  \notag\\
  H_0 &= JN\eta(0)\mathbf{m}^2 +
\frac{1}{2}\sum_{\vk\sigma}\varepsilon_{\vk\sigma} +
\sum_{\vk\alpha\beta} \Delta_{\vk\alpha\beta}^\dag
b_{\vk\alpha\beta}.
\end{align}
Here, $\vk$ is the electron momentum and we have introduced
\begin{equation}
\varepsilon_{\vk\sigma} = \varepsilon_{\vk} -
\sigma\zeta_z,\; \sigma=\uparrow,\downarrow=\pm 1.
\end{equation}
Furthermore, $J$ is a spin coupling constant, $\eta(\vk)$ is a geometrical structure
factor which for $\vk=0$ reduces to the 
number of nearest lattice neighbors $\eta(0)$, $\mathbf{m}=\{m_x,m_y,m_z\}$ is
the magnetization vector, while $\Delta_{\vk\alpha\beta}$ is the
superconducting order parameter and $b_{\vk\alpha\beta} = \langle
c_{-\vk\beta}c_{\vk\alpha}\rangle$ denotes the two-particle operator
expectation value. The ferromagnetic order parameters are given by

\begin{equation}
\zeta = 2J\eta(0)(m_x-\i m_y),\; \zeta_z = 2J\eta(0)m_z.
\end{equation}
 The interesting physics of the FMSC/FMSC
junction lies in the matrix $\hat{{\cal{A}}}_{\vk}$ to be given below.
Above, we used a basis 

\begin{equation}
  \psi_{\vk} =
  (c_{\vk\uparrow}~c_{\vk\downarrow}~c_{-\vk\uparrow}^\dag
  ~c_{\mathbf{-k}\downarrow}^\dag)^{\text{T}},
\end{equation}
where $c_{\vk\sigma}$ ($c_{\vk\sigma}^\dag$) are annihilation
(creation) fermion operators. Note that we have not incorporated any
spin-orbit coupling of the type
$(\mathbf{E}\times\mathbf{p})\cdot\pauliN$ in the Hamiltonian
described in Eq. (\ref{eq:1}) such that spatial inversion symmetry is
not broken, \ie we consider centrosymmetric FMSC.

\par Consider now the matrix
\begin{equation}\label{eq:A}
  \hat{{\cal{A}}}_\vk = -\frac{1}{2}
  \begin{pmatrix}
    -\varepsilon_{\vk\uparrow} & \zeta & \Delta_{\vk\uparrow\uparrow} & \Delta_{\vk\uparrow\downarrow} \\
    \zeta^\dag & -\varepsilon_{\vk\downarrow} &  \Delta_{\vk\downarrow\uparrow} & \Delta_{\vk\downarrow\downarrow} \\
    \Delta^\dag_{\vk\uparrow\uparrow} &  \Delta^\dag_{\vk\downarrow\uparrow} & \varepsilon_{\vk\uparrow} & -\zeta^\dag \\
    \Delta^\dag_{\vk\uparrow\downarrow} & \Delta^\dag_{\vk\downarrow\downarrow} & -\zeta & \varepsilon_{\vk\downarrow}
  \end{pmatrix},
\end{equation}
which is valid for a FMSC with arbitrary magnetization. As explained
in the previous sections, we will study in detail tunneling between
non-unitary ESP FMSC, \ie
$\Delta_{\vk\uparrow\downarrow}=\Delta_{\vk\downarrow\uparrow}=0$,
$\zeta= 0$ in Eq. (\ref{eq:A}). We take the quantization axis on each
side of the junction to coincide with the magnetization direction. One
then needs to include the Wigner $d$-function \cite{wigner1931}
denoted by $\hat{{\cal{D}}}^{(j)}_{\sigma'\sigma}(\vartheta)$ with $j=1/2$
to account for the fact that a $\uparrow$ spin on one side of the
junction is not the same as a $\uparrow$ spin on the other side of the
junction, since the magnetization vectors point can point in different
directions. The angle $\vartheta$ is consequently defined by

\begin{equation}
  \mathbf{m}_\text{R}\cdot\mathbf{m}_\text{L} = m_\text{R}m_\text{L}\cos(\vartheta),\; m_i = |\mathbf{m}_i|.
\end{equation}
Specifically, we have that
\begin{equation}
  \wdf{} = \begin{pmatrix}
    \cos(\vartheta/2) & -\sin(\vartheta/2) \\
    \sin(\vartheta/2) & \cos(\vartheta/2) 
  \end{pmatrix}
\end{equation}
such that a spin-rotated fermion operator is given by 
\begin{equation}
  \widetilde{d}_{\vp\sigma} = \sum_{\sigma'}\wdf{\sigma'\sigma}d_{\vp\sigma'}.
\end{equation}
The tunneling Hamiltonian then reads
\begin{align}
  \label{eq:tunnelham}
  H_\mathrm{T} &= \sum_{\vk\vp\sigma\sigma'}\wdf{\sigma'\sigma} \Big(
  T_{\vk\vp}^{\vphantom{\dagger}}\cc{\vk\sigma}\da{\vp\sigma'}
  + T_{\vk\vp}^{\vphantom{\dagger}\ast}\dc{\vp\sigma'}\ca{\vk\sigma}\Big),
\end{align}
where we neglect the possibility of spin-flips in the tunneling
process. Note that we distinguish between fermion operators on the
right and left side of the junction corresponding to $c_{\vk\sigma}$
and $d_{\vp\sigma}$, respectively. Demanding that $H_\text{T}$ is
invariant under time reversal ${\cal{K}}$, one finds that the
condition ${\cal{K}}^{-1}H_\text{T}{\cal{K}} = H_\text{T}$ with
\begin{align}
  {\cal{K}}^{-1}H_\text{T}{\cal{K}} =& \sum_{\vk\vp\sigma\sigma'}\sigma\sigma' \wdf{\sigma'\sigma}\times\\
  &\Big( T_{\vk\vp}^*c_{-\vk,-\sigma}^\dag d_{-\vp,-\sigma'} + T_{\vk\vp}d_{-\vp,-\sigma'}^\dag c_{-\vk,-\sigma} \Big)\notag
\end{align}
dictates that $T_{\vk\vp} = T_{-\vk,-\vp}^*$. Furthermore, we write the
superconducting order parameters as $ \Delta_{\vk\sigma\sigma} =
|\Delta_{\vk\sigma\sigma}|\e{\i(\theta_\vk +
  \theta^\text{R}_{\sigma\sigma})}$, where R (L) denotes the bulk
superconducting phase on the right (left) side of the junction while
$\theta_\vk$ is a general (complex) internal phase factor originating
from the specific form of the gap in $\vk$-space that ensures odd
symmetry under inversion of momentum, \ie $\theta_\vk = \theta_{-\vk}
+ \pi$.

\par For our system, Eq. (\ref{eq:1}) takes the form
\begin{align}
  H_\text{FMSC} = H_0 + H_A,\; H_A =
  \sum_{\vk\sigma}\phi_{\vk\sigma}^\dag \hat{A}_{\vk\sigma} \phi_{\vk\sigma},
\end{align}
where we have block-diagonalized
$\hat{{\cal{A}}}_\vk$ and chosen a convenient basis $\phi_{\vk\sigma}^\dag =
(c_{\vk\sigma}^\dag, c_{-\vk\sigma})$, with the definition
\begin{equation}
  \hat{A}_{\vk\sigma} = -\frac{1}{2}
  \begin{pmatrix}
    -\varepsilon_{\vk\sigma} & \Delta_{\vk\sigma\sigma}\\
    \Delta_{\vk\sigma\sigma}^\dag & \varepsilon_{\vk\sigma}
  \end{pmatrix}.
\end{equation}
This Hamiltonian is diagonalized by a $2\times 2$ spin generalized
unitary matrix $\hat{U}_{\vk\sigma}$, so that the superconducting sector is
expressed in the diagonal basis
\begin{equation}
  \tilde{\phi}_{\vk\sigma}^\dagger =
  \phi_{\vk\sigma}^\dagger \hat{U}_{\vk\sigma} \equiv (\gc{\vk\sigma},
  \ga{-\vk\sigma}).
\end{equation}
Thus, $H_A =\sum_{\vk\sigma}\tilde{\phi}_{\vk\sigma}^\dagger
\hat{\tilde{A}}_{\vk\sigma} \tilde{\phi}_{\vk\sigma}$, in which

\begin{align}\label{eq:eigenvalues}
  \hat{\tilde{A}}_{\vk\sigma} &= \hat{U}_{\vk\sigma}\hat{A}_{\vk\sigma}\hat{U}_{\vk\sigma}^{-1}
  = \diag(\widetilde{E}_{\vk\sigma}, -\widetilde{E}_{\vk\sigma})/2,\notag\\
  \widetilde{E}_{\vk\sigma} &= 
  \sqrt{\varepsilon_{\vk\sigma}^2+\abs{\Delta_{\vk\sigma\sigma}}^2}.
\end{align}
The explicit expression for $\hat{U}_{\vk\sigma}$ is

\begin{align}\label{eq:U}
  \hat{U}_{\vk\sigma} &= N_{\vk\sigma} 
  \begin{pmatrix}
    1 & \frac{\Delta_{\vk\sigma\sigma}}{\varepsilon_{\vk\sigma}+\widetilde{E}_{\vk\sigma}} \\
    -\frac{\Delta_{\vk\sigma\sigma}^*}{\varepsilon_{\vk\sigma}+\widetilde{E}_{\vk\sigma}} & 1
  \end{pmatrix},\notag\\
  N_{\vk\sigma} &= \frac{\varepsilon_{\vk\sigma}+\widetilde{E}_{\vk\sigma}}{\sqrt{(\varepsilon_{\vk\sigma}+\widetilde{E}_{\vk\sigma})^2 + |\Delta_{\vk\sigma\sigma}|^2}}.
\end{align}
We now proceed to investigate the tunneling currents that can arise
across a junction of two such FMSC.

\subsection{Tunneling formalism}\label{sec:tunneling}
Although the treatment in this section is fairly standard, it comes
with certain extension to the standard cases due to the coexistence of
two simultaneously broken symmetries. Thus, for completeness, we
present it here.

\par In order to find the spin- and charge-current over the junction,
we define the generalized number operator~\footnote{Note that
  $N_{\alpha\beta}$ reduces to the number operator when we sum over
  equal spins, \ie $N=\sum_\sigma N_{\sigma\sigma}$.} by
$N_{\alpha\beta}= \sum_\vk\cc{\vk\alpha}\ca{\vk\beta}$.  Consider now
the transport operator
\begin{align}\label{eq:trans1}
  \dot{N}_{\alpha\beta} &= \im[H_\text{T},N_{\alpha\beta}] \notag\\
  &= -\im\sum_{\vk\vp\sigma}[
  \wdf{\sigma\beta}T_{\vk\vp}\cc{\vk\alpha}\da{\vp\sigma}-\wdf{\sigma\alpha}T_{\vk\vp}^\ast\dc{\vp\sigma}\ca{\vk\beta}].
\end{align}
We now write $H = H' + H_\text{T}$ where $H' = H_\text{L}+H_\text{R}$
and $H_i = K_i + \mu_i N_{i}$, $i$ = L, R, where $\mu_i$ is the
chemical potential on side $i$ and $N_{i}$ is the number operator. In
the interaction picture, the time-dependence of
$\dot{N}_{\alpha\beta}$ is then governed by
\begin{equation}
  \dot{N}_{\alpha\beta}(t) = \e{\i H't}\dot{N}_{\alpha\beta} \e{-\i H't},
\end{equation}
while the time-dependence of the fermion operators reads
\begin{equation}
  c_{\vk\sigma}(t) = \e{\i K_\text{R}t}c_{\vk\sigma}\e{-\i K_\text{R}t}.
\end{equation}
Effectively, one can write
\begin{equation}
  K_\text{R} = H_0 + \sum_{\vk\sigma} E_{\vk\sigma}\gamma_{\vk\sigma}^\dag \gamma_{\vk\sigma},
\end{equation}
where the chemical potential is now included in the quasi-particle
excitation energies $E_{\vk\sigma}$ according to
\begin{equation}
  E_{\vk\sigma} = \sqrt{\xi_{\vk\sigma}^2 + |\Delta_{\vk\sigma\sigma}|^2}
\end{equation}
with $\xi_{\vk\sigma} = \varepsilon_{\vk\sigma} - \mu_\text{R}$, and
correspondingly for the left side.  Consequently, we are able to write
down
\begin{align}
  \label{eq:transopip}
  \begin{split}
    \dot{N}_{\alpha\beta}(t)=-\im\sum_{\vk\vp\sigma}\Big(&
    \wdf{\sigma\beta}T_{\vk\vp}\cc{\vk\alpha}(t)\da{\vp\sigma}(t) 
    \e{-\im teV} 
   \\-&\wdf{\sigma\alpha}T_{\vk\vp}^\ast\dc{\vp\sigma}(t)\ca{\vk\beta}(t) 
    \e{\im teV}\Big),\!
  \end{split}
\end{align}
where $eV\equiv \mu_\text{L}-\mu_\text{R}$ is the externally applied
potential. Within linear response theory, we can identify a general
current
\begin{equation}\label{eq:eqgen}
  \mathbf{I}(t) = \sum_{\alpha\beta} \hat{\boldsymbol{\tau}}_{\alpha\beta}\langle\dot{N}_{\alpha\beta}(t)\rangle,\; \hat{\boldsymbol{\tau}}=(-e\hat{1},\pauliN),
\end{equation}
such that the charge-current is $I^\text{C}(t) = I_0(t)$ while the
spin-current reads $\mathbf{I}^\text{S}(t) = (I_1(t),I_2(t),I_3(t))$.
In Eq. (\ref{eq:eqgen}), $\hat{1}$ denotes the 2$\times$2 identity matrix.
Explicitely, we have
\begin{align}\label{eq:currents}
  I^\text{C}(t) &= I^\text{C}_\text{sp}(t) + I^\text{C}_\text{tp}(t) = -e\sum_{\alpha}\langle \dot{N}_{\alpha\alpha}(t) \rangle\notag\\
  \mathbf{I}^\mathrm{S}(t) &=\mathbf{I}^\text{S}_\text{sp}(t) + \mathbf{I}^\text{S}_\text{tp}(t) = \sum_{\alpha\beta}\pauli\mean{\dot{N}_{\alpha\beta}(t)},
\end{align}
where the subscripts sp and tp denote the single-particle and two-particle
contribution to the currents, respectively. As recently pointed out by
the authors of Ref.~\onlinecite{shi2006}, defining a spin-current is
not as straight-forward as defining a charge-current. Specifically,
the conventional definition of a spin-current given as spin multiplied
with velocity suffers from severe flaws in systems where spin is not a
conserved quantity. In this paper, we define the spin-current across
the junction as $\mathbf{I}^\text{S}(t) = \langle
\text{d}\mathbf{S}(t)/\text{d}t \rangle$ where
$\mathrm{d}\mathbf{S}/\mathrm{d}t = \i[H_\mathrm{T},\mathbf{S}]$. It
is then clear that the concept of a spin-current in this context
refers to the rate at which the spin-vector $\mathbf{S}$ on one side
of the junction changes \textit{as a result of tunneling across the
  junction}. The spatial components of $\mathbf{I}^\text{S}$ are
defined with respect to the corresponding quantization axis. In this
way, we avoid non-physical interpretations of the spin-current in
terms of real spin transport as we only calculate the contribution to
$\mathrm{d}\mathbf{S}/\mathrm{d}t$ from the tunneling Hamiltonian
\textit{instead} of the entire Hamiltonian $H$. Should we have chosen
the latter approach, one would in general run the risk of obtaining a non-zero
spin-current due to \eg local spin-flip processes which are obviously
not relevant in terms of real spin transport across the junction. However, in our system such spin-flip processes are absent.

\par
The tunneling currents are calculated in the linear response regime by
using the Kubo formula,
\begin{equation}\label{eq:Kubo}
  \mean{\dot{N}_{\alpha\beta}(t)} = -\im\int_{-\infty}^{t}\dif t'
  \mean{[\dot{N}_{\alpha\beta}(t),H_\mathrm{T}(t')]},
\end{equation}
where the right hand side is the statistical expectation value in the
unperturbed quantum state, \ie when the two subsystems are not
coupled. This expression includes both single-particle and
two-particle contributions to the current.  Details of the
calculations are found in Sec. \ref{app:fs}.

\par  
We now consider the cases of an A2- and A1-phase at zero external
potential, giving special attention to the charge-current and
$\hat{\mathbf{z}}$-component of the spin-current in the Josephson
channel.

\subsection{Two-particle currents}
For an A2-phase in the case of zero externally applied voltage
($eV=0$), Eqs. (\ref{eq:currents}) and (\ref{eq:GenMat}) generates a 
quasiparticle interference term $I_\text{qi}$, in addition to a term $I_\text{J}$
identified as the Josephson current. Thus, the total two-particle currents of charge and spin can be written as $I^\text{C(S)}_{\text{tp}(,z)} = I^\text{C(S)}_{\text{qi}(,z)} + I^\text{C(S)}_{\text{J}(,z)}$ where 
\begin{align}
I^\text{C(S)}_{\text{qi}(,z)} &= \sum_{\vk\vp} I^\text{C(S)}(\theta^\text{L}_{\sigma\sigma} - \theta^\text{R}_{\alpha\alpha}, \Delta\theta_{\vp\vk}),\notag\\
I^\text{C(S)}_{\text{J}(,z)} &= \sum_{\vk\vp} I^\text{C(S)}(\Delta\theta_{\vp\vk}, \theta^\text{L}_{\sigma\sigma} - \theta^\text{R}_{\alpha\alpha})
\end{align}
with the definitions
\begin{align}
  \label{eq:currentsA2}
  I^\text{C}(\phi_1,\phi_2) = 
  \frac{e}{2}\sum_{\sigma\alpha} &[1+\sigma\alpha\cos(\vartheta)]|T_{\vk\vp}|^2
  \frac{|\Delta_{\vk\alpha\alpha}|
    |\Delta_{\vp\sigma\sigma}|}{E_{\vk\alpha}E_{\vp\sigma}}\notag\\
  &\times\cos(\phi_1)\sin(\phi_2)
  F_{\vk\vp\alpha\sigma},\notag\\
  I^\text{S}(\phi_1,\phi_2) = 
  -\frac{1}{2}\sum_{\sigma\alpha} \alpha &[1+\sigma\alpha\cos(\vartheta)]|T_{\vk\vp}|^2
  \frac{|\Delta_{\vk\alpha\alpha}|
    |\Delta_{\vp\sigma\sigma}|}{E_{\vk\alpha}E_{\vp\sigma}}\notag\\
  &\times\cos(\phi_1)\sin(\phi_2)
  F_{\vk\vp\alpha\sigma},
\end{align}
where we have introduced $\Delta\theta_{\vp\vk} \equiv \theta_\vp - \theta_\vk$ and

\begin{align}
  F_{\vk\vp\alpha\sigma} 
  &= \sum_\pm \frac{f(\pm E_{\vk\alpha})-f(E_{\vp\sigma})}{E_{\vk\alpha} \mp E_{\vp\sigma}}.
\end{align}
Above, $f(x)$ is the Fermi distribution. Thus, we have found a two-particle current, for both spin and charge,
that can be tuned in a well-defined manner by adjusting the relative
orientation $\vartheta$ of the magnetization vectors \footnote{For corresponding results in spin-singlet superconductors with helimagnetic order, see Refs.~\onlinecite{kulic2005,eremin2006}.}. We will discuss
the detection of such an effect later in this paper. Note that the $\vk$-dependent symmetry factor $\theta_\vk$ enters the above expressions, thus giving rise to an extra contribution to the two-particle current besides the ordinary Josephson effect. This is due to the fact that we included it in the SC gaps as a factor $\e{\i\theta_\vk}$ which in general is complex. However, this specific form may for certain models, depending on the Fermi surface in question, be reduced to a real function, \ie $\e{\i\theta_\vk} \to \cos\theta_\vk$, in which case the quasi-particle interference term becomes zero. Hence, in most of the remaining discussion we will focus on the Josephson part of the two-particle current.

\par
The A1-phase with only one SC order
parameter $\Delta_{\vk\alpha\alpha}$,
$\alpha\in\{\uparrow,\downarrow\}$ also corresponds to a non-unitary
state $\mathbf{d}_\vk$ according to Eq. (\ref{eq:Cooperpairspin}), and is
thus compatible with coexistence of FM and SC. In this case, we
readily see that Eq.  (\ref{eq:currentsA2}) reduces to
\begin{align}
\label{eq:currentsA1}
  \begin{split}
    I^\text{C}_\text{tp} &= e\cos^2(\vartheta/2)X_{\alpha} \\
    I^\text{S}_{\text{tp},z} &= -\alpha\cos^2(\vartheta/2)X_{\alpha}
  \end{split}
  \qquad
  \alpha\in\{\uparrow,\downarrow\}
\end{align}
where we have defined the quantity
\begin{align}
X_{\alpha} &= \sum_{\vk\vp}|T_{\vk\vp}|^2\frac{|\Delta_{\vk\alpha\alpha}\Delta_{\vp\alpha\alpha}|}{E_{\vk\alpha}E_{\vp\alpha}}F_{\vk\vp\alpha\alpha}\notag\\
  &\times[\sin\Delta\theta_{\alpha\alpha}\cos\Delta\theta_{\vp\vk} + \cos\Delta\theta_{\alpha\alpha}\sin\Delta\theta_{\vp\vk}]
\end{align}
with $\Delta\theta_{\alpha\alpha} \equiv
\theta^\text{L}_{\alpha\alpha}-\theta^\text{R}_{\alpha\alpha}$, and
$\Delta_{\vk\alpha\alpha}$ is the surviving order parameter. As
expected, the spin-current changes sign depending on whether it is the
$\Delta_{\vk\uparrow\uparrow}$ or $\Delta_{\vk\downarrow\downarrow}$
order parameter that is present.

\par
For collinear magnetization $(\vartheta=0)$, an ordinary Josephson
effect occurs with the superconducting phase difference as the driving
force. Interestingly, one is able to tune both the spin- and
charge-current to zero in the A1-phase when $\mathbf{m}_\text{L}
\parallel -\mathbf{m}_\text{R}$ ($\vartheta=\pi$). It follows from Eq.
(\ref{eq:currentsA1}) that the spin- and charge-current only differ by
a constant pre-factor
\begin{equation}
  I^\text{C}_\text{tp}/I^\text{S}_{\text{tp},z} = -\alpha e,\quad \alpha=\pm1.
\end{equation} 
It is then reasonable to draw the conclusion that we are dealing with
a completely \textit{spin-polarized current} such that both
$I^\text{C}_\text{tp}$ and $I^\text{S}_{\text{tp},z}$ must vanish
simultaneously at $\vartheta=\pi$. 

\par
Another result that can be extracted from Eqs. (\ref{eq:currentsA2})
and (\ref{eq:currentsA1}) is a persistent non-zero DC spin-Josephson
current even if the magnetizations on each side of the junction are of
equal magnitude and collinear ($\vartheta=0$). This is quite different
from the spin-Josephson effect recently considered in ferromagnetic
metal junctions ~\cite{nogueira2004b}. In that case, a twist in the
magnetization across the junction is required to drive the
spin-Josephson effect.

\par
Note that in the common approximation $T_{\vk\vp} = T$, \ie the tunneling probability is independent of the electron magnitude and direction of electron momentum, the two-particle current predicted above is identically equal to zero. Of course, such a crude approximation does not correspond to the correct physical picture (see \eg Ref.~\onlinecite{bruder1995}), and in general one cannot neglect the directional dependence of the tunneling matrix element. This demonstrates that we are dealing with a more subtle effect than what could be unveiled when applying the approximation of a constant tunneling matrix element.

\par
An interesting situation arises in the case of zero externally applied
voltage \textit{and} identical superconductors on each side of the
junction with SC phase differences
$\Delta\theta_{\sigma\sigma} = 0$. In this case, we find that $I^\mathrm{C}_\mathrm{J} =0$ while
\begin{align}\label{eq:finalspincrazy}
  I^\text{S}_{\text{J},z} = -2\sum_{\vk\vp} &|T_{\vk\vp}|^2 \sin^2(\vartheta/2) |\Delta_{\vk\uparrow\uparrow}\Delta_{\vp\downarrow\downarrow}|F_{\vk\vp\uparrow\downarrow}\notag\\
  &\times \sin(\theta_{\downarrow}^\text{L}-\theta_{\uparrow}^\text{R})/(E_{\vk\uparrow}E_{\vp\downarrow}).
\end{align}
when $eV=0$, $\Delta\theta_{\sigma\sigma}=0$. Thus, we have found a
dissipationless spin-current in the two-particle channel without an
externally applied voltage \textit{and} without a SC phase difference.
This effect is present as long as $\vartheta$ is not 0 or $\pi$,
corresponding to parallel or anti-parallel magnetization on each side
of the junction. It is seen from Eq. (\ref{eq:finalspincrazy}) that
the spin-current is driven by an interband phase difference on each
side of the junction. A necessary condition for this effect to occur
is that no inter-band Josephson coupling is present, \ie electrons in
the two energy-bands $E_{\vk\uparrow}$ and $E_{\vk\downarrow}$ do not
communicate with each other. To understand why a Josephson coupling
would destroy the above effect, consider the free energy density for a
$p$-wave FMSC first proposed in Ref.~\onlinecite{walker2002}, given by
\begin{equation}\label{eq:joscop}
  {\cal{F}} = {\cal{F}}' - \lambda_\text{J}\cos(\theta_{\uparrow\uparrow}-\theta_{\downarrow\downarrow})
\end{equation}
in the presence of a Josephson coupling. In Eq. (\ref{eq:joscop}),
$\lambda_\text{J}$ determines the strength of the interaction while
${\cal{F}}'$ contains the SC and FM contribution to the free energy
density in addition to the coupling terms between the SC and FM order
parameters. Consequently, the phase difference
$\theta_{\uparrow\uparrow}-\theta_{\downarrow\downarrow}$ is locked to
0 or $\pi$ in order to minimize ${\cal{F}}$, depending on
sgn($\lambda_\text{J}$). Considering Eq. (\ref{eq:finalspincrazy}), we
see that $I^\text{S}_{\text{J},z}=0$ in this case, since the argument
of the last sine is zero. Mechanisms that would induce a Josephson
coupling include magnetic impurities causing inelastic spin-flip
scattering between the energy-bands and spin-orbit coupling. Recently,
the authors of Ref.~\onlinecite{shi2006} proposed that $p$-wave SC
arising out of a FM metal state could be explained by the Berry
curvature field that is present in ferromagnets with spin-orbit
coupling. It is clear that in the case where spin-orbit coupling is
included in the problem, spin-flip scattering processes occur between
the energy bands such that the $\uparrow$ and $\downarrow$ spins can
not be considered as two independent species any more. The SC phases
will then be locked to each other with a relative phase of $0$ or
$\pi$. However, note that in the general case, Eq.
(\ref{eq:currentsA2}) produces a non-zero charge- and spin-current
even if the spin-up and spin-down phases are locked to each other.

\subsection{Single-particle currents}
In the single-particle channel, we find that the charge and
spin-currents read
\begin{align}\label{eq:spcurrent}
  I^\text{C}_{\text{sp}} &= -e\sum_\alpha \langle \dot{N}_{\alpha\alpha}(t) \rangle_\text{sp} \notag\\
  I^\text{S}_{\text{sp},z} &= \sum_{\alpha} \alpha  \langle \dot{N}_{\alpha\alpha}(t) \rangle_\text{sp},
\end{align}
as seen from Eq. (\ref{eq:currents}). From Eq. (\ref{eq:N1}), we then
extract the proper expectation value, which is found to be
\begin{widetext}
  \begin{align}
    \langle \dot{N}_{\alpha\alpha}(t) \rangle_\text{sp} &= 4\pi\sum_{\vk\vp\sigma} [1+\sigma\alpha\cos(\vartheta)]|T_{\vk\vp\alpha}|^2 N_{\vk\alpha}^2N_{\vp\sigma}^2\notag\\
    &\times\Bigg[ [f(E_{\vk\alpha})-f(E_{\vp\sigma})]\Big( \delta(-eV+E_{\vk\alpha}-E_{\vp\sigma}) - \frac{|\Delta_{\vk\alpha\alpha}\Delta_{\vp\sigma\sigma}|^2}{(\xi_{\vk\alpha}+E_{\vk\alpha})^2(\xi_{\vp\sigma}+E_{\vp\sigma})^2   }\delta(-eV-E_{\vp\sigma}+E_{\vk\alpha}) \Big)\notag\\
    &+ [1-f(E_{\vk\alpha})-f(E_{\vp\sigma})]
    \Big( \frac{|\Delta_{\vk\alpha\alpha}|^2}{(\xi_{\vk\alpha}+E_{\vk\alpha})^2 }\delta(-eV-E_{\vk\alpha}-E_{\vp\sigma}) - \frac{|\Delta_{\vp\sigma\sigma}|^2}{(\xi_{\vp\sigma}+E_{\vp\sigma})^2 }\delta(-eV+E_{\vk\alpha}+E_{\vp\sigma})\Big)  \Bigg].
  \end{align}
\end{widetext}

The currents in Eq. (\ref{eq:spcurrent}) are thus seen to require an
applied voltage in order to flow in the tunneling junction. Clearly, this is because the Cooper pairs need to be split up in
order for a single-particle current to exist, such that both spin- and
charge-currents vanish at $eV=0$.

\par
In Ref.~\onlinecite{nogueira2004b}, the presence of a persistent
spin-current in the single-particle channel for FM/FM junctions with a
twist in magnetization across the junction was predicted. For
consistency, our results must confirm
this prediction for the single-particle current in the limit
where SC is lost, \ie $\Delta_{\vk\sigma\sigma}\to0$. Note that the $\hat{\mathbf{z}}$-direction in
Ref.~\onlinecite{nogueira2004b} corresponds to a vector in our local
$xy$-plane since the present quantization axis lies parallel with the
magnetization direction. Upon calculating the $\mathbf{x}$- and
$\mathbf{y}$-components of the single-particle spin-current for our
system in the limit where SC is lost, \ie $\Delta_{\vk\sigma\sigma}\to
0$, a persistent spin Josephson-like current proportional to
$\sin(\vartheta)$ is identified. More precisely,
\begin{align}\label{eq:nogcompare}
  \ve{I}^\text{S}_{\text{sp}}(t) = 
  2\sum_{\vk\vp}\sum_{\alpha\beta\sigma}&
  \wdf{\sigma\alpha}\wdf{\sigma\beta}\abs{\Tkp{}}^2\notag\\
  &\times \Imag{\big\{\pauli[\beta\alpha]}\Lambda_{\beta\sigma}^{1,1}(-eV)\big\}
\end{align}
when $\Delta_{\vq\sigma\sigma}=0$ (see Appendix for details). In agreement with
Ref.~\onlinecite{nogueira2004b}, the component of the spin-current
parallel to $\ve{m}_\text{L}\times\ve{m}_\text{R}$ is seen to vanish
for $\vartheta=\{0,\pi\}$ at $eV=0$.

\section{Ferromagnets with spin-orbit coupling}\label{sec:fmsoc}

\subsection{Coexistence of ferromagnetism and spin-orbit coupling}
In a system where time-reversal and spatial inversion symmetry are
simultaneously broken, it is clear that spins are heavily affected by
these properties. There is currently much focus on ferromagnetic
semiconductors where spin-orbit coupling plays a crucial role with
regard to transport properties \cite{dietl2002, matsukura2002}. In
fact, there has in recent years been much progress in the
semiconductor research community where the spin-Hall effect in
particular has received much attention \cite{engel2006}. With the
discovery \cite{ohno1992} of hole-mediated ferromagnetic order in
(In,Mn)As, extensive research on III-V host materials was triggered.
Moreover, it is clear that properties such as ferromagnetic transition
temperatures in excess of 100 K \cite{matsukura1998} and long
spin-coherence times \cite{kikkawa1999} in GaAs have strongly
contributed to opening up a vista plethora for information processing
and storage technologies in these new magnetic mediums
\cite{jungwirth2002}.

Generally, spin-orbit coupling (SOC) can be roughly divided into two
categories -- \emph{intrinsic} and \emph{extrinsic}. Intrinsic SOC is
found in materials with a non-centrosymmetric crystal symmetry, \ie
where inversion symmetry is broken,
whereas extrinsic SOC is due to asymmetries caused by impurities,
local confinements of electrons or externally applied electrical
fields.

In the present paper, we investigate the tunneling current of spin
between two ferromagnetic metals with spin-orbit coupling induced by
an external electric field. This way, we will have two externally
controllable parameters; the magnetization $\ve{m}$ and the electrical field $\ve{E}$. The case of tunneling between 
two noncentrosymmetric superconductors with significant 
spin-orbit coupling, but no ferromagnetism, has previously 
been considered in  Ref. \onlinecite{borkje2006}.

\subsection{The Hamiltonian}
\begin{figure}[!ht]
  \centering
  \resizebox{0.49\textwidth}{!}{
    \includegraphics{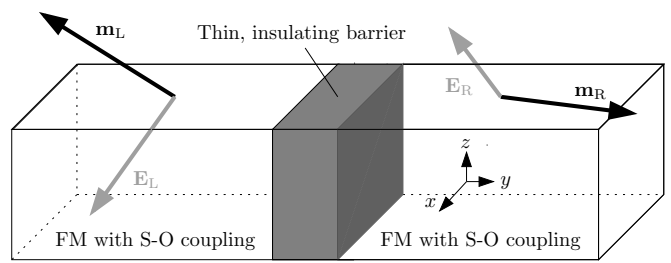}}
  \caption{Our model consisting of two ferromagnetic metals with
    spin-orbit coupling separated by a thin insulating barrier. The
    magnetization $\mathbf{m}$ and electrical field $\mathbf{E}$ are
    allowed to point in any direction so that our results are
    generally valid, while special cases such as planar magnetization
    etc. are easily obtained by applying the proper limits to the
    general expressions.}
  \label{fig:setup}
\end{figure}

Our system consists of two Heisenberg ferromagnets with substantial
spin-orbit coupling, separated by a thin insulating barrier which is
assumed to be spin-inactive. This is shown in Fig. \ref{fig:setup}. We
now operate with only one quantization axis, such that a proper
tunneling Hamiltonian for this purpose is 
\begin{equation}
H_\text{T}=
\sum_{\vk\vp\sigma}(T_{\vk\vp} c_{\vk\sigma}^\dag d_{\vp\sigma} +
\text{h.c.}),
\end{equation}
 where $\{c^\dag_{\vk\sigma},c_{\vk\sigma}\}$ and
$\{d^\dag_{\vk\sigma},d_{\vk\sigma}\}$
 are creation and annihilation
operators for an electron with momentum $\vk$ and spin $\sigma$ on the
right and left side of the junction, respectively, while $T_{\vk\vp}$
is the spin-independent tunneling matrix element.
In $\vk$-space, the Hamiltonian describing the ferromagnetism reads
\begin{equation}
  H_\text{FM} = \sum_{\vk\sigma}\varepsilon_{\vk} c_{\vk\sigma}^\dag c_{\vk\sigma}- JN\sum_{\vk}\eta(\vk) \mathbf{S}_\vk\cdot\mathbf{S}_{-\vk}
\end{equation}
in which $\varepsilon_{\vk}$ is the kinetic energy of the electrons,
$J$ is the ferromagnetic coupling constant, $N$ is the number of
particles in the system, while $\mathbf{S}_\vk = (1/2) \sum_{\alpha\beta}
c_{\vk\alpha}^\dag\pauli c_{\vk\beta}$ is the spin operator. As we
later adopt the mean-field approximation, $\mathbf{m}=(m_x,m_y,m_z)$
will denote the magnetization of the system.
\par
The spin-orbit interactions are accounted for by a Rashba
Hamiltonian
\begin{equation}
H_\text{S-O} = -\sum_{\vk}
\varphi_\vk^\dag  [\xi(\nabla V\times\vk)\cdot\hat{\boldsymbol{\sigma}}]\varphi_\vk,
\end{equation}
where $\varphi_\vk = [c_{\vk\uparrow}, c_{\vk\downarrow}]^\text{T}$,
$\mathbf{E}=-\nabla V$ is the electrical field felt by the electrons
and $\hat{\boldsymbol{\sigma}}=(\hat{\sigma}_1,\hat{\sigma}_2,\hat{\sigma}_3)$ in which
$\hat{\sigma}_i$ are Pauli matrices, while the parameter $\xi$ is
material-dependent. From now on, the notation
$\xi(\mathbf{E}\times\vk)\equiv\mathbf{B}_\vk =
(B_{\vk,x},B_{\vk,y},B_{\vk,z})$ will be used. In general, the
electromagnetic potential $V$ consists of two parts $V_\text{int}$ and
$V_\text{ext}$ (see \eg Ref.~\onlinecite{engel2006} for a detailed
discussion of the spin-orbit Hamiltonian). The crystal potential of
the material is represented by $V_\text{int}$, and only gives rise to
a spin-orbit coupling if inversion symmetry is broken in the crystal
structure.  Asymmetries such as impurities and local confinements of
electrons are included in $V_\text{ext}$, as well as any external
electrical field.  Note that any lack of crystal inversion symmetry
results in a so-called Dresselhaus term in the Hamiltonian, which is
present in the absence of any impurities and confinement potentials.
In the following, we focus on the spin-orbit coupling resulting from
$V_\text{ext}$, thus considering any symmetry-breaking electrical
field that arises from charged impurities or which is applied
externally. In the case where the crystal structure does not respect inversion
symmetry, a Dresselhaus term \cite{dresselhaus1955} can be 
easily included in the Hamiltonian by performing the substitution
\begin{equation}
(\mathbf{E}\times\vk)\cdot\hat{\boldsymbol{\sigma}} \to
[(\mathbf{E}\times\vk) + {\cal{D}}(\vk)]\cdot\hat{\boldsymbol{\sigma}},
\end{equation}
where ${\cal{D}}(\vk)$ = $-{\cal{D}}(-\vk).$

\par
We now proceed to calculate the spin-current that is generated across
the junction as a result of tunneling. Note that in our model, the
magnetization vector and electrical field are allowed to point in
arbitrary directions. In this way, the obtained result for the
spin-current will be generally valid and special cases, \eg thin
films, are easily obtained by taking the appropriate limits in the
final result. It should be mentioned that the effective 
magnetic field from the spin-orbit interactions might influence the direction
of the magnetization in the ferromagnet. This is, however, not the main focus of our work, and we leave this question open for study. Our emphasis in the present paper concerns the derivation of general results onto which specific restrictions may be applied as they seem appropriate.

\par
In the mean-field approximation, the Hamiltonian for the right side of
the junction can be written as $H= H_\text{FM}+H_\text{S-O}$, which in
a compact form yields
\begin{equation}\label{eq:H1}
H_\mathrm{R} =H_0 + \sum_{\vk} \varphi_\vk^\dag 
\begin{pmatrix}
\varepsilon_{\vk\uparrow} & -\zeta_\text{R} + \bk \\
-\zeta^\dag_\text{R}  + B_{\vk,+} & \varepsilon_{\vk\downarrow} 
\end{pmatrix}
\varphi_\vk,
\end{equation}
where $\varepsilon_{\vk\sigma}\equiv\varepsilon_\vk - \sigma
(\zeta_{z,\text{R}} - B_{\vk,z})$ and $H_0$ is an irrelevant constant.
The FM order parameters are $\zeta_\text{R} =
2J\eta(0)(m_{\text{R},x}-\i m_{\text{R},y})$ and $\zeta_{z,\text{R}} =
2J\eta(0)m_{\text{R},z}$ and $B_{\vk,\pm} \equiv B_{\vk,x}\pm\i B_{\vk,y}$.  For
convenience, we from now on write $\zeta=|\zeta|\e{\i\phi}$ and $B_{\vk,\pm} =
|B_{\vk,\pm}|\e{\mp\i\chi_\vk}$. The Hamiltonian for the left side of the
junction is obtained from Eq. (\ref{eq:H1}) simply by the doing the
replacements $\vk\to\vp$ and $\text{R}\to\text{L}$.

\subsection{Tunneling formalism}
\par
In order to obtain the expressions for the spin- and charge- tunneling
currents, it is necessary to calculate the Green functions. These are
given by the matrix 
\begin{equation}
{\hat{\cal{G}}}_\vk(\i\omega_n) =
(-\i\omega_n\hat{1}+\hat{\A})^{-1},
\end{equation} where $\hat{\A}$ is the matrix in Eq. (\ref{eq:H1}).
Explicitly, we have that
\begin{equation}
  \hat{{\cal{G}}}_\vk(\i\omega_n)=\begin{pmatrix}
    G_\vk^{\uparrow\uparrow}(\i\omega_n)& F_\vk^{\downarrow\uparrow}(\i\omega_n)\\
    F_\vk^{\uparrow\downarrow}(\i\omega_n) & G_\vk^{\downarrow\downarrow}(\i\omega_n)\\
  \end{pmatrix}.
\end{equation}
Above, $\omega_n = 2(n+1)\pi/\beta, n=0,1,2\ldots$ is the fermionic
Matsubara frequency and $\beta$ denotes inverse temperature.
Introducing 
\begin{equation}
X_{\vk}(\i\omega_n) =
(\varepsilon_{\vk\uparrow}-\i\omega_n)(\varepsilon_{\vk\downarrow}-\i\omega_n)
- |\zeta_\mathrm{R}-\bk|^2,
\end{equation}
the normal and anomalous Green functions are
\begin{align}\label{eq:green}
  G_\vk^{\sigma\sigma}(\i\omega_n) &= (\varepsilon_{\vk,-\sigma} -
  \i\omega_n)/X_{\vk}(\i\omega_n),\;\notag\\
 F_\vk^{\downarrow\uparrow}(\i \omega_n) &=
 F_\vk^{\uparrow\downarrow,\dag}(\i\omega_n) =
 (\zeta_\text{R}-\bk)/X_{\vk}(\i\omega_n).
\end{align} 

\par 
The expression for $\mathbf{I}^\text{S}(t)$ is established by first
considering the generalized number operator $N_{\alpha\beta} =
\sum_\vk c_{\vk\alpha}^\dag c_{\vk\beta}$. This operator changes with
time due to tunneling according to $\dot N_{\alpha\beta} =
\i[H_\text{T},N_{\alpha\beta}]$, which in the interaction picture
representation becomes $\dot N_{\alpha\beta}(t) =
-\i\sum_{\vk\vp}(T_{\vk\vp} c_{\vk\alpha}^\dag d_{\vp\beta}\e{\-\i
  teV} - \text{h.c.}).$ The voltage drop across the junction is given
by the difference in chemical potential on each side, \ie $eV =
\mu_\text{R}-\mu_\text{L}$. In the linear response regime, the spin-current across the junction is
\begin{equation}
  \mathbf{I}^\text{S}(t) = \frac{1}{2}\sum_{\alpha\beta}\pauli \langle \dot N_{\alpha\beta} (t) \rangle,
\end{equation}
where the expectation value of the time derivative of the transport
operator is calculated by means of the
Kubo formula Eq. (\ref{eq:Kubo}). Details will be given in Sec.
\ref{app:fmsoc}.

\subsection{Single-particle currents}
At $eV=0$, it is readily seen from the discussion in Sec.
\ref{app:fmsoc} that the charge-current vanishes.  Consider now the
$z$-component of the spin-current in particular, which can be written
as $I^\text{S}_z = \Im\text{m}\{\Phi(-eV)\}$. The Matsubara function
$\Phi(-eV)$ is found by performing analytical continuation $\i \wnu \to
-eV + \i 0^+$ on $\widetilde{\Phi}(\i\wnu)$, where
\begin{align}\label{eq:mat}
  \widetilde{\Phi}(\i\wnu)=\frac{1}{\beta}\sum_{\i \omega_m, \vk\vp}
  \sum_\sigma \sigma 
  \Big(  &G_\vk^{\sigma\sigma}(\i\omega_m) G^{\sigma\sigma}_\vp(\i\omega_m -\i \wnu) \notag\\
  + &F^{-\sigma,\sigma}_\vk(
  \i\omega_m)F^{\sigma,-\sigma}_\vp(\i\omega_m -\i \wnu) \Big).
\end{align}
Here, $\wnu = 2\nu\pi/\beta$, $\nu=0,1,2\ldots$ is the bosonic
Matsubara frequency.  Inserting the Green functions from Eq.
(\ref{eq:green}) into Eq. (\ref{eq:mat}), one finds that a persistent
spin-current is established across the tunneling junction. For zero
applied voltage, we obtain
\begin{subequations}\label{eq:finalspin}
  \begin{align}
    I^\text{S}_z &= \sum_{\vk\vp}  \frac{|T_{\vk\vp}|^2J_{\vk\vp}}{2\gamma_\vk\gamma_\vp}\Big[|\zeta_\text{R}\zeta_\text{L}|\sin\Delta\phi + |\bk\bp|\sin\Delta\chi_{\vk\vp}\notag\\
    &-|\bk\zeta_\text{L}|\sin(\chi_\vk-\phi_\text{L}) - |\bp\zeta_\text{R}|\sin(\phi_\text{R}-\chi_\vp)\Big], \\
    J_{\vk\vp} &= \sum_{\substack{\alpha=\pm\\\beta=\pm}}
    \alpha\beta\Bigg[\frac{n(\varepsilon_{\vk}+\alpha\gamma_\vk)-n(\varepsilon_\vp+\beta\gamma_\vp)}{(\varepsilon_\vk+\alpha\gamma_\vk)-(\varepsilon_\vp+\beta\gamma_\vp)}\Bigg].
  \end{align}
\end{subequations}
In Eqs. (\ref{eq:finalspin}), $\Delta\chi_{\vk\vp} \equiv \chi_\vk
- \chi_\vp$, $\Delta\phi \equiv \phi_\text{R}-\phi_\text{L}$, while
\begin{equation}\label{eq:gamma}
  \gamma_\vk^2= (\zeta_{z,\text{R}} - B_{\vk,z})^2 +
  |\zeta_\text{R} - \bk|^2
\end{equation}
and $n(\varepsilon)$ denotes the Fermi distribution. In the above
expressions, we have implicitly associated the right side R with the
momentum label $\vk$ and L with $\vp$ for more concise notation, such
that \eg $B_{\vk,z} \equiv B_{\vk,z}^\text{R}$. Defining $\zeta_i =
2J\eta(0)m_i$, we see that Eq. (\ref{eq:gamma}) can be written as
\begin{equation}\label{eq:gammavec}
  \gamma_\vk = |\boldsymbol{\zeta}_\text{R} - \mathbf{B}_\vk|.
\end{equation}

\par
The spin-current described in Eq. (\ref{eq:finalspin}) can be
controlled by adjusting the relative orientation of the magnetization
vectors on each side of the junction, \ie $\Delta\phi$, and also
responds to a change in direction of the applied electric fields. The
presence of an external magnetic field $\mathbf{H}_i$ would control
the orientation of the internal magnetization $\mathbf{m}_i$.
Alternatively, one may also use exchange biasing to an
anti-ferromagnet in order to lock the magnetization direction.
Consequently, the spin-current can be manipulated by the external
control parameters $\{\mathbf{H}_i,\mathbf{E}_i\}$ in a well-defined
manner. This observation is highly suggestive in terms of novel
nanotechnological devices.

\par
We stress that Eq. (\ref{eq:finalspin}) is \textit{non-zero} in the
general case, since $\gamma_\vk \neq -\gamma_{-\vk}$ and
$\chi_{-\vk} = \chi_{\vk}+\pi$. Moreover, Eq. (\ref{eq:finalspin})
is valid for any orientation of both $\mathbf{m}$ and $\mathbf{E}$ on
each side of the junction, and a number of interesting special cases
can now easily be considered simply by applying the appropriate limits
to this general expression.

\subsection{Special limits}
\begin{figure}[!ht]
\centering
\resizebox{0.49\textwidth}{!}{
\includegraphics{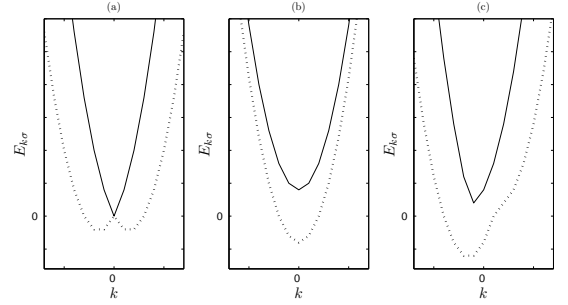}}
\caption{Schematic illustration of the energy-bands for (a) a system
  with spin-orbit coupling, (b) a system with ferromagnetic ordering,
  and (c) a system exhibiting both of the aforementioned
  properties. The dotted line corresponds to quasi-particles with $\sigma=\downarrow$, while the full drawn line designates $\sigma=\uparrow$. Since the density of states
  $N^\sigma(E_{k\sigma})$ is proportional to
  $(\partial E_{k\sigma}/\partial k)^{-1}$, we see that a
  difference between $N^\uparrow( E_{k\sigma})$ and
  $N^\downarrow( E_{k\sigma})$ is zero at Fermi level in (a), while the
  density of states differ for the $\uparrow$- and
  $\downarrow$-populations in (b) and (c). Thus, a persistent
  spin-current will only occur for tunneling between systems
  corresponding to (b) and (c).}
\label{fig:energy}
\end{figure}

Consider first the limit where ferromagnetism is absent, such that the
tunneling occurs between two bulk materials with spin-orbit coupling.
Applying $\mathbf{m}\to 0$ to Eq. (\ref{eq:finalspin}), it is readily
seen that the spin-current vanishes for any orientation of the
electrical fields. Intuitively, one can understand this by considering
the band structure of the quasi-particles with energy $E_{\vk\sigma} =
\varepsilon_\vk + \sigma\gamma_\vk$ and the corresponding density of
states $N(E_{\vk\sigma})$ when only spin-orbit coupling is present, as
shown in Fig. \ref{fig:energy}. Since the density of states is equal
for $\uparrow$ and $\downarrow$ spins \footnote{Note that the index
  $\alpha$ on the quasi-particles does not denote the physical spin of
  electrons, but is rather to be considered as some unspecified
  helicity-index. The usage of the word ``spin'' in this context then
  refers to this helicity.}, one type of spin is not preferred
compared to the other with regard to tunneling, resulting in a net
spin-current of zero. Formally, the vanishing of the spin-current can
be understood by replacing the momentum summation with integration
over energy, \ie $\sum_{\vk\vp} \to \int\int \mathrm{d} E_\text{R}
\mathrm{d} E_\text{L} N_\mathrm{R}(E_\text{R})N_\mathrm{L}
(E_\text{L})$.  When $\mathbf{m}\to 0$, Eq.  (\ref{eq:finalspin})
dictates that
\begin{align}\label{eq:integral}
  I^\mathrm{S}_z  \sim \sum_{\substack{\alpha=\pm\\\beta=\pm}}
  \alpha\beta \int \int&
  \mathrm{d} E_{\mathrm{R},\alpha}\mathrm{d} E_{\mathrm{L},\beta}
  N^\alpha_\mathrm{R}(E_{\mathrm{R},\alpha})N^\beta_\mathrm{L}(E_{\mathrm{L},\beta})\notag\\
  &\times
  \Bigg[\frac{n(E_{\mathrm{R},\alpha})-n(E_{\mathrm{L},\beta})}{E_{\mathrm{R},\alpha}
    - E_{\mathrm{L},\beta}}\Bigg].
\end{align}
Since the density of states for the $\uparrow$- and
$\downarrow$-populations are equal in the individual subsystems, \ie
$N^\uparrow(E)=N^\downarrow(E)\equiv
N(E)$, the integrand of Eq. (\ref{eq:integral}) becomes
spin-independent such that the summation over $\alpha$ and $\beta$
yields zero. Thus, no spin-current will exist at $eV=0$ over a
tunneling barrier separating two systems with spin-orbit coupling
alone. In the general case where both ferromagnetism and spin-orbit
coupling are present, the density of states at, say, Fermi level are
different, leading to a persistent spin-current across the junction
due to the difference between $N^\uparrow(E)$ and
$N^\downarrow(E)$.

\par
\begin{figure}[!ht]
\centering
\resizebox{0.46\textwidth}{!}{
\includegraphics{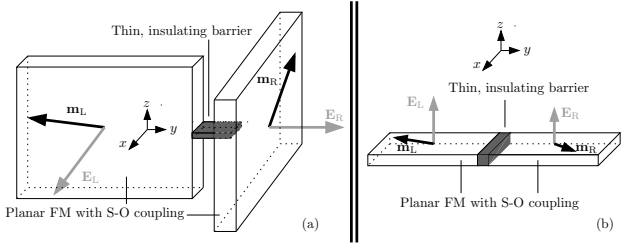}}
\caption{Tunneling between planar ferromagnets in the presence of externally applied electrical fields $\mathbf{E}_\text{L}$ and $\mathbf{E}_\text{R}$ that destroy inversion symmetry and induce a spin-orbit coupling.}
\label{fig:specialcase}
\end{figure}

We now consider a special case where the bulk structures indicated in
Fig. \ref{fig:setup} are reduced to two thin-film ferromagnets in the
presence of electrical fields that are perpendicular to each other,
say $\mathbf{E}_\text{L} = (E_\text{L},0,0)$ and $\mathbf{E}_\text{R}
= (0,E_\text{R},0)$, as shown in Fig. \ref{fig:specialcase}(a) and
(b). In this case, we have chosen an in-plane magnetization for each
of the thin-films. Solving specifically for Fig.
\ref{fig:specialcase}(a), it is seen that $\mathbf{m}_\text{L} =
(0,m_{\text{L},y}, m_{\text{L},z})$ and $\mathbf{m}_\text{R} =
(m_{\text{R},x},0, m_{\text{R},z})$. Furthermore, assume that the
electrons are restricted from moving in the ``thin'' dimension, \ie $\vp
= (0,p_y,p_z)$ and $\vk = (k_x,0,k_z)$. In this case, Eq.
(\ref{eq:finalspin}) reduces to the form
\begin{equation}
  I^\text{S}_z = I_0\text{sgn}(m_{\text{L},y}) + \sum_{\vk\vp} I_{1,\vk\vp}\text{sgn}(p_z),
\end{equation}
where the constants above are 
\begin{align}\label{eq:I0}
  I_0 &= \sum_{\vk\vp} \frac{|T_{\vk\vp}|^2 J_{\vk\vp}(|\zeta_\mathrm{R}\zeta_\mathrm{L}|-E_\mathrm{R}|k_z\zeta_\mathrm{L}|)}{2|\boldsymbol{\zeta}_\mathrm{R}+\mathbf{B}_\vk||\boldsymbol{\zeta}_\mathrm{L} + \mathbf{B}_\vp|},\notag\\
  I_{1,\vk\vp} &= \frac{|T_{\vk\vp}|^2J_{\vk\vp}E_\mathrm{L}(E_\mathrm{R}|k_zp_z| - |p_z\zeta_\mathrm{R}|)}{2|\boldsymbol{\zeta}_\mathrm{R}+\mathbf{B}_\vk||\boldsymbol{\zeta}_\mathrm{L} + \mathbf{B}_\vp|},
\end{align}
with
\begin{align}
  \boldsymbol{\zeta}_\mathrm{L} &= 2J\eta(0)(0,m_{\mathrm{L},y},m_{\mathrm{L},z}) &\mathbf{B}_\vp = \xi_\mathrm{L}E_\mathrm{L}(0,-p_z,p_y)& \notag\\
  \boldsymbol{\zeta}_\mathrm{R} &= 2J\eta(0)(m_{\mathrm{R},x},0,m_{\mathrm{R},z}) &\mathbf{B}_\vk = \xi_\mathrm{R}E_\mathrm{R}(k_y,0,-k_x)& \notag\\
\end{align}
such that $I_{1,\vk\vp} \neq I_{1,-\vk,-\vp}$. Likewise, for the setup sketched in Fig. \ref{fig:specialcase}(b), one obtains
\begin{align}
  I^\mathrm{S}_z &= \sum_{\vk\vp}\frac{|T_{\vk\vp}|^2 J_{\vk\vp}}{2|\zeta_\mathrm{R} + E_\mathrm{R}(k_y-\i k_x)|^2|\zeta_\mathrm{L}+E_\mathrm{L}(p_y-\i p_x)|^2}\notag\\
  &\times\Big[|\zeta_\mathrm{R}\zeta_\mathrm{L}|\sin\Delta\phi + E_\mathrm{R}E_\mathrm{L}(k_y^2+k_x^2)(p_y^2+ p_x^2)|\sin\Delta\chi_{\vk\vp}\notag\\
  &\hspace{0.4in}-E_\mathrm{R}|\zeta_\mathrm{L}|(k_y^2+k_x^2)\sin(\chi_k-\phi_\mathrm{L}) \notag\\
  &\hspace{0.4in}- \mathrm{E_L}|\zeta_\mathrm{R}|(p_y^2+
  p_x^2)\sin(\phi_\mathrm{R}-\chi_\vp)\Big],
\end{align}
where $\chi_{\vq}$ obeys
\begin{equation}
\tan \chi_{\vq} = -\frac{q_x}{q_y},\; \vq = \vk,\vp.
\end{equation}

\par From these observations, we can draw the following conclusions:
whereas the spin-current is zero for the system in Fig.
\ref{fig:specialcase}(a) and (b) if only spin-orbit coupling is
considered, it is non-zero when only ferromagnetism is taken into
account. However, in the general case where both ferromagnetism and
spin-orbit coupling are included, an \textit{additional term} in the
spin-current is induced compared to the pure ferromagnetic case.
Accordingly, there is an interplay between the magnetic order and the
Rashba-interaction that produces a spin-current which is more than
just the sum of the individual contributions.

\section{Discussion of results}\label{sec:discuss}
Having presented the general results for tunneling currents between
systems with multiple broken symmetries in the preceeding sections, we
now focus on detection and experimental issues concerning verification
of our predictions.

\par
Consider first the system consisting of two ferromagnetic
superconductors separated by a thin, insulating barrier.  It is
well-known that for tunneling currents flowing between two $s$-wave SC
in the presence of a magnetic field that is perpendicular to the
tunneling direction, the resulting flux threading the junction leads
to a Fraunhofer-like variation in the DC Josephson effect, given by a
multiplicative factor
\begin{equation}
  D_\text{F}(\Phi) = \frac{\sin(\pi\Phi/\Phi_0)}{(\pi\Phi/\Phi_0)}
\end{equation}
in the critical current. Here, $\Phi_0 = \pi\hbar/e$ is the elementary
flux quantum, and $\Phi$ is the total flux threading the junction due
to a magnetic field.  Consequently, the presence of magnetic flux in
the tunneling junction of two $s$-wave SC threatens to nullify the
total Josephson current. In the present case of two $p$-wave FMSC,
this is not an issue since we have assumed uniform coexistence of the
SC and FM order parameters which is plausible for a weak intrinsic
magnetization. The effect of an external magnetic field $\mathbf{H}$
would then simply be to rotate the internal magnetization as dictated
by the term $-\mathbf{H}\cdot\mathbf{m}$ in the free energy
${\cal{F}}$ (see \eg Ref.~\onlinecite{shopova2005}). Thus, there is no
diffraction pattern present for the tunneling-currents between two
non-unitary ESP FMSC, regardless of how the internal magnetization is
oriented. Since the motion of the Cooper-pairs is also restricted by
the thin-film structure, there is no orbital effect from such a
magnetization.

Note that the interplay between ferromagnetism and superconductivity
is manifest in the charge- as well as spin-currents, the former being
readily measurable. Detection of the induced spin-currents would be
challenging, although recent studies suggest feasible methods of
measuring such quantities \cite{malshukov2005}. We comment more on
this later in this section. First, we adress the issue of how boundary effects
affect the order parameters. Studies \cite{ambegaokar1974, buchholtz1981, tanuma2001} have shown that interfaces/surfaces may
have a pair-breaking effect on unconventional SC order parameters. This is highly relevant in tunneling junction 
experiments as in the present case. The suppression of the order parameter is caused by a formation of
so-called midgap surface states (also known as zero-energy states) \cite{hu1994} which occurs for certain orientations of the $\vk$-dependent SC gaps that satisfy a resonance condition. Note that this is not the case for conventional $s$-wave
superconductors since the gap is isotropic in that case. This pair-breaking surface effect was studied specifically for $p$-wave order parameters in Refs.~\onlinecite{ambegaokar1974, buchholtz1981}, and it was found that the component of the order parameter that experiences a sign change under the transformation $k_\perp \to -k_\perp$, where $k_\perp$ is the component of momentum perpendicular to the tunneling junction, was suppressed in the vicinity of the junction. By vicinity of the junction, we here mean a distance comparable to the coherence length, typically of order 1-10 nm. Thus, depending on the explicit form of the superconducting gaps in the FMSC, these could be subject to a reduction close to the junction, which in turn would reduce the magnitude of the Josephson effect we predict. Nevertheless, the latter is nonvanishing in the general case.
\par 
Since the critical Josephson currents depend on the relative
magnetization orientation, one is able to tune these currents in a
well-defined manner by varying $\vartheta$. This can be done by
applying an external magnetic field in the plane of the FMSC. In the
presence of a rotating magnetic moment on either side of the junction,
the Josephson currents will thus vary according to Eq.
(\ref{eq:currentsA2}), which may be cast into the form
$I^\mathrm{C}_\mathrm{J} = I_0 + I_m\cos(\vartheta)$. Depending on the
relative magnitudes of $I_0$ and $I_m$, the sign of the critical
current may change. Note that such a variation of the magnetization
vectors must take place in an adiabatic manner so that the systems can
be considered to be in, or near, equilibrium at all times. Our
predictions can thus be verified by measuring the critical current at
$eV=0$ for different angles $\vartheta$ and compare the results with
our theory.  Recently, it has been reported that a spin-triplet
supercurrent, induced by Josephson tunneling between two $s$-wave
superconductors across a ferromagnetic metallic contact, can be
controlled by varying the magnetization of the ferromagnetic contact
\cite{keizer2006}. Moreover, concerning the spin-Josephson current we
propose, detection of induced spin-currents are challenging, although
recent studies suggest feasible methods of measuring such quantities
\cite{malshukov2005}. Observation of macroscopic spin-currents in
superconductors may also be possible via angle resolved photo-emission
experiments with circularly polarized photons \cite{simon2002}, or in
spin-resolved neutron scattering experiments \cite{hirsch1999}.

\par
We reemphasize that the above ideas should be experimentally
realizable by \eg utilizing various geometries in order to vary the
demagnetization fields.  Alternatively, one may use exchange biasing
to an anti-ferromagnet.  Such techniques of achieving non-collinearity
are routinely used in ferromagnet-normal metal structures
~\cite{bass1999}.

\par
With regard to the predicted DC spin-current in for a system
consisting of two ferromagnetic metals with spin-orbit coupling, we
here suggest how this effect could be probed for in an experimental
setup. For instance, the authors of Ref.~\onlinecite{mohanty2004}
propose a spin-mechanical device which exploits nanomechanical torque
for detection and control of a spin-current. Similarly, a setup
coupling the electron spin to the mechanical motion of a
nanomechanical system is proposed in Ref.~\onlinecite{malshukov2005}.
The latter method employs the strain-induced spin-orbit interaction of
electrons in a narrow gap semiconductor. In Ref.~\onlinecite{sun2004},
it was demonstrated that a steady-state magnetic-moment current, \ie
spin-current, will induce a static electric field. This fact may be
suggestive in terms of detection \cite{meier2003, schutz2003}, and
could be useful to observe the novel effects predicted in this paper.

\section{Summary}\label{sec:summary}

In summary, we have considered supercurrents of spin and charge that
exist in FMSC/FMSC and FMSO/FMSO tunneling junctions. In the former
case, we have found an interplay between the relative magnetization
orientation on each side of the junction and the SC phase difference
when considering tunneling between two non-unitary ESP FMSC with
coexisting and uniform FM and SC order. This interplay is present in
the Josephson channel, offering the opportunity to tune
dissipationless currents of spin and charge in a well-defined manner
by adjusting the relative magnetization orientation on each side of
the junction. As a special case, we considered the case where the SC
phase difference is zero, and found that a dissipationless
spin-current \textit{without} charge-current would be established
across the junction. Suggestions concerning the detection of the
effects we predict have been made.

\par
Moreover, we have derived an expression for a dissipationless
spin-current that arises in the junction between two Heisenberg
ferromagnets with spin-orbit coupling. We have shown that the
spin-current is driven by terms originating from both the
ferromagnetic phase difference, in agreement with the result of
Ref.~\onlinecite{nogueira2004b}, and the presence of spin-orbit
coupling itself. In addition, it was found that the simultaneous
breaking of time-reversal and inversion symmetry fosters an interplay
between ferromagnetism and spin-orbit coupling in the spin-current.
Availing oneself of external magnetic and electric fields, our
expressions show that the spin-current can be tuned in a well-defined
manner. These results are of significance in the field of spintronics
in terms of quantum transport, and offer insight into how the
spin-current behaves for nanostructures exhibiting both ferromagnetism
and spin-orbit coupling.

\section*{Acknowledgments}
The authors acknowledge J.-M. B{\o}rven for his valuable contribution,
and thank A. Brataas, K. B{\o}rkje, and E. K. Dahl for helpful
discussions.  This work was supported by the Norwegian Research
Council Grants No. 157798/432 and No. 158547/431 (NANOMAT), and Grant
No. 167498/V30 (STORFORSK).

\appendix

\section{Details of Matsubara formalism}
\subsection{Ferromagnetic superconductors}\label{app:fs}
Inserting Eq. (\ref{eq:transopip}) into Eq. (\ref{eq:Kubo}), one finds
that
\begin{align}\label{eq:N1}
  \langle \dot{N}_{\alpha\beta}(t)\rangle &= \langle \dot{N}_{\alpha\beta}(t)\rangle_\text{sp} + \langle \dot{N}_{\alpha\beta}(t)\rangle_\text{tp} \notag\\
  &= -\int^t_{-\infty}\dif t \Big[   \langle [M_{\alpha\beta}(t),M^\dag(t')]\rangle\e{-\i eV(t-t')} \notag\\
  &\;\;\;\;\;\;\;- \langle [M_{\beta\alpha}^\dag(t),M(t')]\rangle\e{\i eV(t-t')}            \notag\\
  &\;\;\;\;\;\;\;+ \langle [M_{\alpha\beta}(t),M(t')]\rangle\e{-\i eV(t+t')}\notag\\
  &\;\;\;\;\;\;\; - \langle
  [M_{\beta\alpha}^\dag(t),M^\dag(t')]\rangle\e{\i eV(t+t')} \Big]
\end{align}
where the two first terms in Eq. (\ref{eq:N1}) contribute to the
single-particle current while the two last terms constitute the
Josephson current. Above, we defined
\begin{align}\label{eq:M}
  M_{\alpha\beta}(t) &= \sum_{\vk\vp\sigma} \wdf{\sigma\beta}T_{\vk\vp}c_{\vk\alpha}^\dag(t)d_{\vp\sigma}(t) \notag\\
  M(t) &= \sum_{\vk\vp\sigma\sigma'} \wdf{\sigma'\sigma} T_{\vk\vp}
  c_{\vk\sigma}^\dag(t)d_{\vp\sigma'}(t).
\end{align}
By observing that $\pauli = \pauliDAG$, we can combine Eqs.
(\ref{eq:Kubo})-(\ref{eq:M}) to yield
\begin{align}\label{eq:N2}
  \pauli\mean{\dot{N}_{\alpha\beta}(t)}_\text{sp} &= 2\Im\text{m}\{ \pauli\Phi_{\alpha\beta,\text{sp}}(-eV) \}\notag\\
  \pauli\mean{\dot{N}_{\alpha\beta}(t)}_\text{tp} &= 2\Im\text{m}\{
  \pauli\Phi_{\alpha\beta,\text{J}}(eV)\e{-2\i etV} \}
\end{align}
where the Matsubara functions are obtained by performing analytical
continuation according to
\begin{align}\label{eq:tildemat}
  \Phi_{\alpha\beta,\text{sp}}(-eV) &= \lim_{\i\wnu\to -eV+\i 0^+} \widetilde{\Phi}_{\alpha\beta,\text{sp}}(\i\wnu) \notag\\
  \Phi_{\alpha\beta,\text{J}}(eV) &= \lim_{\i\wnu\to eV+\i 0^+}
  \widetilde{\Phi}_{\alpha\beta,\text{tp}}(\i\wnu),
\end{align}
In Eq. (\ref{eq:tildemat}), $\wnu = 2\pi \nu/\beta,\; \nu=1,2,3\ldots$
is the bosonic Matsubara frequency and

\begin{align}\label{eq:matsubara}
  \widetilde{\Phi}_{\text{sp},\alpha\beta}(\i\wnu) &=
  -\int^\beta_0 \text{d}\tau \e{\i\wnu\tau}
  \sum_{\stackrel{\vk\vp\sigma}
    {_{\vk'\vp'\sigma_1\sigma_2}}}\wdf{\sigma\beta}\wdf{\sigma_1\sigma_2}\times\notag\\
  &T_{\vk\vp}T_{\vk'\vp'}^* \langle\TT\{ c_{\vk\alpha}^\dag(\tau)d_{\vp\sigma}(\tau)d_{\vp'\sigma_1}^\dag(0)c_{\vk'\sigma_2}(0)\} \rangle,\notag\\
  \widetilde{\Phi}_{\text{tp},\alpha\beta}(\i\wnu) &= -\int^\beta_0
  \text{d}\tau \e{\i\wnu\tau} \sum_{\stackrel{\vk\vp\sigma}
    {_{\vk'\vp'\sigma_1\sigma_2}}}\wdf{\sigma\beta}\wdf{\sigma_1\sigma_2}\times\notag\\
  &T_{\vk\vp}T_{\vk'\vp'}\langle\TT\{
  c_{\vk\alpha}^\dag(\tau)d_{\vp\sigma}(\tau)c_{\vk'\sigma_2}^\dag(0)d_{\vp'\sigma_1}(0)\}\rangle.
\end{align}

Here, $\TT$ denotes the time-ordering operator, and
$\beta=1/k_\text{B}T$ is the inverse temperature. Only $\vk'=(-)\vk,
\vp'=(-)\vp$ contributes in the single-particle (two-particle) channel,
while the diagonalized basis $\widetilde{\varphi}_{\vk\sigma}$
dictates that only $\sigma_2=\alpha,\; \sigma_1=\sigma$ contributes in
the spin summation. Making use of the relation
$\widetilde{\phi}_{\vk\sigma}^\dagger = \phi_{\vk\sigma}^\dagger
\hat{U}_{\vk\sigma}$, Eq. (\ref{eq:matsubara}) becomes
\begin{align}\label{eq:mat1}
  \widetilde{\Phi}_{\text{sp},\alpha\beta}(\i\wnu) =& \int^\beta_0 \text{d}\tau \e{\i\wnu\tau} \sum_{\vk\vp\sigma} \wdf{\sigma\beta}\wdf{\sigma\alpha} T_{\vk\vp}T_{\vk\vp}^* \notag\\
  &\times  \langle\TT\{ \Big[  \hat{U}_{11\vk\alpha}^*\gamma_{\vk\alpha}^\dag(\tau) + \hat{U}_{12\vk\alpha}^*\gamma_{-\vk\alpha}(\tau) \Big]\notag\\
  &\times\Big[ \hat{U}_{11\vk\alpha}\gamma_{\vk\alpha}(0) + \hat{U}_{12\vk\alpha}\gamma_{-\vk\alpha}^\dag(0)\Big]\} \rangle\notag\\
  &\times  \langle \TT\Big[ \hat{U}_{11\vp\sigma}\gamma_{\vp\sigma}(\tau) + \hat{U}_{12\vp\sigma}\gamma_{-\vp\sigma}^\dag(\tau) \Big]\notag\\
  &\times \Big[ \hat{U}_{11\vp\sigma}^*\gamma_{\vk\sigma}^\dag(0) +
  \hat{U}_{12\vp\sigma}^*\gamma_{-\vk\sigma}(0) \Big]\} \rangle
\end{align}
\begin{align}
  \widetilde{\Phi}_{\text{tp},\alpha\beta}(\i\wnu) =& -\int^\beta_0 \text{d}\tau \e{\i\wnu\tau} \sum_{\vk\vp\sigma} \wdf{\sigma\beta} \wdf{\sigma\alpha} T_{\vk\vp}T_{-\vk,-\vp} \notag\\
  &\times\langle \TT\{\Big[\hat{U}_{11\vk\alpha}^*\gamma_{\vk\sigma}^\dag(\tau) + \hat{U}_{12\vk\alpha}^*\gamma_{-\vk\sigma}(\tau)\Big]\notag\\
  &\times\Big[\hat{U}_{21\vk\alpha}\gamma_{\vk\sigma}(0) + \hat{U}_{22\vk\alpha}\gamma_{-\vk\sigma}^\dag(0) \Big]\}\rangle\notag\\
  &\times\langle\TT \Big[ \hat{U}_{21\vp\sigma}^*\gamma_{\vp\sigma}^\dag(0) + \hat{U}_{22\vp\sigma}^*\gamma_{-\vp\sigma}(0) \Big]\notag\\
  &\times\Big[ \hat{U}_{11\vp\sigma}\gamma_{\vp\sigma}(\tau) +
  \hat{U}_{12\vp\sigma}\gamma_{-\vp\sigma}^\dag(\tau)\Big]\}\rangle
\end{align}
Since our diagonalized Hamiltonian has the form of a free-electron
gas, \ie

\begin{align}
  H_\text{FMSC} = \widetilde{H}_0 + \sum_{\vk\sigma} E_{\vk\sigma}
  \gamma_{\vk\sigma}^\dag\gamma_{\vk\sigma}
\end{align}
with $\widetilde{H}_0 = H_0 - (E_{\vk\uparrow}+E_{\vk\downarrow})$,
the product of the new fermion operators
$\widetilde{\varphi}_{\vk\sigma}$ in Eq. (\ref{eq:mat1}) yield
unperturbed Green's functions according to

\begin{equation}\label{eq:Green}
  G_\alpha(\vk,\tau-\tau') = \langle \TT\{c_{\vp\alpha}^\dag(\tau')c_{\vk\alpha}(\tau)\} \rangle
\end{equation}
We then Fourier-transform Eq. (\ref{eq:Green}) into
\begin{equation}
  G_\alpha(\vk,\tau) = \frac{1}{\beta} \sum_{\wm} \e{-\i\wm\tau}G_\alpha(\vp,\i\wm),
\end{equation}
where $\wm = (2m+1)\pi/\beta$, $m=1,2,3\ldots$ is a
fermionic Matsubara frequency. The frequency summation over $m$ is
evaluated by contour integration as in \eg Ref.~\onlinecite{mahan2002}
to yield the result

\begin{align}
  \frac{1}{\beta}\sum_m G_\alpha(\vk,\i\wm)G_\sigma(\vp,\i\wnu+&\i\wm) = \frac{f(E_{\vk\alpha})-f(E_{\vp\sigma})}{\i\wnu + E_{\vk\alpha} - E_{\vp\sigma}}\notag\\
  \frac{1}{\beta}\sum_mG_\alpha(\vk,\i\wm)G_\sigma(\vp,\i\wnu-&\i\wm)=\frac{f(E_{\vp\sigma})-f(-E_{\vk\alpha})}{\i\wnu
    - E_{\vk\alpha} - E_{\vp\sigma}},
\end{align}
where $f(E)=1-f(-E)=1/(1+\e{\beta E})$ is the Fermi distribution. It
is then a matter of straight-forward calculations to obtain the result
  \begin{align}
    \label{eq:GenMatsp}
    \Phi_{\text{sp},\alpha\beta}(-eV)& = \sum_{\vk\vp\sigma} \wdf{\sigma\beta}\wdf{\sigma\alpha} T_{\vk\vp}T_{\vk\vp}^*N_{\vk\alpha}^2N_{\vp\sigma}^2\notag\\
    \Bigg[& \frac{|\Delta_{\vk\alpha\alpha}\Delta_{\vp\sigma\sigma}|^2 \Lambda_{\vk\vp\sigma\alpha}^{-1,-1}(-eV)}{(\xi_{\vk\alpha}+E_{\vk\alpha})(\xi_{\vp\sigma}+E_{\vp\sigma})} +
    \Lambda_{\vk\vp\sigma\alpha}^{1,1}(-eV) \notag\\
    &+ \frac{|\Delta_{\vp\sigma\sigma}|^2\Lambda_{\vk\vp\sigma\alpha}^{-1,1}(-eV)}{\xi_{\vp\sigma}+E_{\vp\sigma}} \notag\\
    &+\frac{|\Delta_{\vk\alpha\alpha}|^2\Lambda_{\vk\vp\sigma\alpha}^{1,-1}(-eV)}{\xi_{\vk\alpha}+E_{\vk\alpha}}
     \Bigg]
\end{align}
\begin{align}\label{eq:GenMat}
  \Phi_{\text{tp},\alpha\beta}(eV) =& -\sum_{\vk\vp\sigma}
  \wdf{\sigma\beta} \wdf{\sigma\alpha} T_{\vk\vp}T_{-\vk,-\vp}\notag\\
&\times\frac{\Delta_{\vk\alpha\alpha}^\ast\Delta_{\vp\sigma\sigma}}{
    4E_{\vk\alpha}E_{\vp\sigma}}
  \sum_{\substack{\x=\pm 1\\\y=\pm 1}}
  \Lambda_{\vk\vp\sigma\alpha}^{\x\y}(eV),
\end{align}
where $\Lambda_{\vk\vp\sigma\alpha}^{\x\y}(eV)$ is obtained by
performing analytical continuation $\im\wnu \to eV + \im 0^+$ on
\begin{equation}
  \widetilde{\Lambda}_{\vk\vp\sigma\alpha}^{\x\y}(\im\wnu)
  =\frac{\x[f(E_{\vk\alpha}) - f(\x\y E_{\vp\sigma})]}{\im\wnu + \y
    E_{\vk\alpha}-\x E_{\vp\sigma}};\quad \x,\y=\pm 1.
\end{equation}

We also provide the details of the persistent spin-supercurrent for
$\Delta_{\sigma\sigma}=0$. Writing the Josephson current Eq.
(\ref{eq:currentsA2}) out explicitly, one has that
$I^\text{C}_\text{J} = e I^+$ and $I^\text{S}_\text{J} = -I^-$ where

\begin{align}\label{eq:spincrazy}
  I^\pm =& \sum_{\vk\vp} |T_{\vk\vp}|^2 \Bigg[ \cos^2(\vartheta/2) \frac{|\Delta_{\vk\uparrow\uparrow}\Delta_{\vp\uparrow\uparrow}|}{E_{\vk\uparrow}E_{\vp\uparrow}}\sin\Delta\theta_{\uparrow\uparrow} F_{\vk\vp\uparrow\uparrow} \notag\\
  &+ \sin^2(\vartheta/2) \frac{|\Delta_{\vk\uparrow\uparrow}\Delta_{\vp\downarrow\downarrow}|}{E_{\vk\uparrow}E_{\vp\downarrow}}\sin(\theta^\text{L}_{\downarrow\downarrow} - \theta^\text{R}_{\uparrow\uparrow}) F_{\vk\vp\uparrow\downarrow} \notag\\
  &\pm \sin^2(\vartheta/2) \frac{|\Delta_{\vk\downarrow\downarrow}\Delta_{\vp\uparrow\uparrow}|}{E_{\vk\downarrow}E_{\vp\uparrow}} \sin(\theta^\text{L}_{\uparrow\uparrow} - \theta^\text{R}_{\downarrow\downarrow})  F_{\vk\vp\downarrow\uparrow} \notag\\
  &\pm \cos^2(\vartheta/2)
  \frac{|\Delta_{\vk\downarrow\downarrow}\Delta_{\vp\downarrow\downarrow}|}{E_{\vk\downarrow}E_{\vp\downarrow}}\sin\Delta\theta_{\downarrow\downarrow}
  F_{\vk\vp\downarrow\downarrow}\Bigg].
\end{align}
The first and fourth term above vanish when
$\Delta\theta_{\sigma\sigma}=0$. By observing that
$F_{\vk\vp\uparrow\downarrow} = F_{\vp\vk\downarrow\uparrow}$, we are
then able to re-write Eq. (\ref{eq:spincrazy}) as
\begin{align}
  I^\pm =& \sum_{\vk\vp} |T_{\vk\vp}|^2 \sin^2(\vartheta/2) \frac{|\Delta_{\vk\uparrow}\Delta_{\vp\downarrow}|}{E_{\vk\uparrow}E_{\vp\downarrow}}F_{\vk\vp\uparrow\downarrow}\notag\\
  &\times \Big[\sin(\theta_{\downarrow}^\text{L} - \theta_{\uparrow}^\text{R}) \pm \sin(\theta_{\uparrow}^\text{L} - \theta_{\downarrow}^\text{R}) \Big] \notag\\
  =&  e\sum_{\vk\vp} |T_{\vk\vp}|^2 \sin^2(\vartheta/2) \frac{|\Delta_{\vk\uparrow}\Delta_{\vp\downarrow}|}{E_{\vk\uparrow}E_{\vp\downarrow}}F_{\vk\vp\uparrow\downarrow}\notag\\
  &\times \Big[\sin\Big( (\theta_{\downarrow}^\text{L}\mp\theta_{\downarrow}^\text{R} - \theta_{\uparrow}^\text{R}\pm\theta_{\uparrow}^\text{L} )/2   \Big)\notag\\
  &\times \cos\Big(
  (\theta_{\downarrow}^\text{L}\pm\theta_{\downarrow}^\text{R} -
  \theta_{\uparrow}^\text{R}\mp\theta_{\uparrow}^\text{L} )/2 \Big)
  \Big].
\end{align}
It is clear that the argument of the sine gives 0 for the upper sign,
such that $I^\text{C}_\text{J}=0$. But for the lower sign, the
argument of the cosine is equal to 0, such that Eq.
(\ref{eq:finalspincrazy}) is obtained.

\subsection{Ferromagnets with spin-orbit coupling}\label{app:fmsoc}

The spin-current across the junction can be written as
\begin{align}
  \I &= \Im\mathrm{m}\{\bPhi(-eV)\},\notag\\
  \bPhi(-eV) &= \lim_{\i\wnu\to -eV+\i
    0^+}\widetilde{\bPhi}(\i\wnu),
\end{align}
where we have defined the Matsubara function
\begin{align}\label{eq:matt1}
  \widetilde{\bPhi}(\i\wnu) = \sum_{\vk\vp\alpha\beta\sigma} &|T_{\vk\vp}|^2\pauli\int^\beta_0\mathrm{d}\tau\e{\i\wnu\tau}\notag\\
  \times&\langle\T\{c_{\vk\sigma}(0)c_{\vk\alpha}^\dag(\tau)\}\rangle\langle\T\{d_{\vp\beta}(\tau)d_{\vp\sigma}^\dag(0)\}\rangle.
\end{align}
In Eq. (\ref{eq:matt1}), we defined the time-ordering operator $\T$
while $\beta$ in the upper integration limit is inverse temperature
and $\wnu = 2n\pi/\beta, n=0,1,2,\ldots$ is a bosonic Matsubara
frequency. From the definition of the spin-generalized Green's
function
\begin{equation}
  G^{\alpha\beta}_\vk(\tau-\tau') = -\langle\T\{c_{\vk\alpha}(\tau)c_{\vk\beta}^\dag(\tau')\}\rangle,
\end{equation}
Eq. (\ref{eq:matt1}) can be written out explicitely to yield
\begin{widetext}
  \begin{align}\label{eq:mat2}
    \widetilde{\bPhi}(\i\wnu) = \frac{1}{\beta}
    \sum_{\vk\vp,m}|T_{\vk\vp}|^2
    \Bigg[&\pauliN_{\uparrow\uparrow}\Big(G^{\uparrow\uparrow}_\vk(\i\wm)G^{\uparrow\uparrow}_\vp(\i\wm
    - \i\wnu) +
    G^{\downarrow\uparrow}_\vk(\i\wm)G^{\uparrow\downarrow}_\vp(\i\wm
    - \i\wnu)\Big)
    +\pauliN_{\uparrow\downarrow}\Big(G^{\uparrow\uparrow}_\vk(\i\wm)G^{\downarrow\uparrow}_\vp(\i\wm - \i\wnu) \notag\\
    &+
    G^{\downarrow\uparrow}_\vk(\i\wm)G^{\downarrow\downarrow}_\vp(\i\wm
    - \i\wnu)\Big)
    +\pauliN_{\downarrow\uparrow}\Big(G^{\uparrow\downarrow}_\vk(\i\wm)G^{\uparrow\uparrow}_\vp(\i\wm - \i\wnu) + G^{\downarrow\downarrow}_\vk(\i\wm)G^{\uparrow\downarrow}_\vp(\i\wm - \i\wnu)\Big) \notag\\
    &+\pauliN_{\downarrow\downarrow}\Big(G^{\uparrow\downarrow}_\vk(\i\wm)G^{\downarrow\uparrow}_\vp(\i\wm
    - \i\wnu) +
    G^{\downarrow\downarrow}_\vk(\i\wm)G^{\downarrow\downarrow}_\vp(\i\wm
    - \i\wnu)\Big)\Bigg].
  \end{align}
\end{widetext}
We made use of the Fourier-transformations
\begin{align}
  G^{\alpha\beta}_\vk(\i\wm) &= \int^\beta_0 \mathrm{d}\tau\e{\i\wm}G^{\alpha\beta}_\vk(\tau),\notag\\
  G^{\alpha\beta}_\vk(\tau) &= \frac{1}{\beta} \sum_m
  \e{-\i\wm\tau}G^{\alpha\beta}_\vk(\i\wm)
\end{align}
in writing down Eq. (\ref{eq:mat2}), where $\wm = 2(m+1)\pi/\beta,
m=0,1,2,\ldots$ is a fermionic Matsubara frequency.  Having written
down the full expression for the Matsubara function in Eq.
(\ref{eq:mat2}), one can now easily distinguish between components of
the spin-current. For instance, only $\pauliN_{\alpha\alpha}$ will
contribute to the $\hat{\mathbf{z}}$-component of $\I$, and the
corresponding terms can be read out from Eq. (\ref{eq:mat2}).  From
the present Green functions in Eq. (\ref{eq:green}), it is obvious
that three types of frequency summations must be performed, namely
\begin{align}
  J_{\vk\vp,r} = \frac{1}{\beta} &\sum_m\Bigg[ 
  \frac{\wm^r}{[(\varepsilon_{\vk\uparrow}-\i\wm)(\varepsilon_{\vk\downarrow}-\i\wm)-y_\vk^2]}\notag\\
  \times&\frac{1}{[(\varepsilon_{\vp\uparrow}-\i\wm+\i\wnu)(\varepsilon_{\vp\downarrow}-\i\wm+\i\wnu)-y_\vp^2]}\Bigg],
\end{align}
with $r$ is an integer. Performing the summation over $m$ using residue calculus, one finds that
\begin{align}
  J_{\vk\vp,r} &= \sum_{\substack{\alpha=\pm\\\beta=\pm}}
  \frac{\alpha\beta}{4y_\vk
    y_\vp}\Bigg[\frac{\psi^r_{\vk\alpha}n(\psi_{\vk\alpha})-(\i\wnu+\psi_{\vp\beta})^rn(\psi_{\vp\beta})}{-\i\wnu+\psi_{\vk\alpha}-\psi_{\vp\beta}}\Bigg]
\end{align}
with the definition $\psi_{\vk\alpha} \equiv \varepsilon_{\vk}+\alpha
y_\vk$. Separating the general expression Eq. (\ref{eq:mat2}) into its
spatial components $\widetilde{\bPhi} =
(\widetilde{\Phi}_x,\widetilde{\Phi}_y,\widetilde{\Phi}_z)$, the components of the spin-current can be extracted according to
$I_i^\text{S} = \Im\mathrm{m}\{\Phi_i(-eV)\}$, $i=x,y,z.$ Note that the charge-current in this model, which vanishes for $eV=0$, is obtained by the performing the replacement $\hat{\boldsymbol{\sigma}}_{\alpha\beta} \to \hat{1}_{\alpha\beta}$, where $\hat{1}$ is the 2$\times$2 unit matrix. We find
that
\begin{widetext}
  \begin{align}\label{eq:mat4}
    \widetilde{\Phi}_x(\i\wnu) &= \sum_{\vk\vp} \frac{|T_{\vk\vp}|^2}{4\gamma_\vk\gamma_\vp}\Bigg[J_{\vk\vp,0}\Big( \varepsilon_{\vk\downarrow}(\zeta_\text{L}-\bp) + (\varepsilon_{\vp\uparrow}+\i\wnu)(\zeta_\text{R}-\bk) + (\varepsilon_{\vp\downarrow}+\i\wnu)(\zeta_\text{R}^\dag - B_{\vk,+}) + \varepsilon_{\vk\uparrow}(\zeta_\text{L}^\dag - \bp) \Big) \notag\\
    &\hspace{0.8in} -J_{\vk\vp,1}\Big( (\zeta_\text{L}-\bp) + (\zeta_\text{R}-\bk) + (\zeta_\text{R}^\dag - B_{\vk,+}) + (\zeta_\text{L}^\dag -B_{\vp,+})  \Big)\Bigg],\notag\\
    \widetilde{\Phi}_y(\i\wnu) &= \sum_{\vk\vp} \i\frac{|T_{\vk\vp}|^2}{4\gamma_\vk\gamma_\vp}\Bigg[J_{\vk\vp,0}\Big( -\varepsilon_{\vk\downarrow}(\zeta_\text{L}-\bp) - (\varepsilon_{\vp\uparrow}+\i\wnu)(\zeta_\text{R}-\bk) + (\varepsilon_{\vp\downarrow}+\i\wnu)(\zeta_\text{R}^\dag - B_{\vk,+}) + \varepsilon_{\vk\uparrow}(\zeta_\text{L}^\dag - \bp) \Big) \notag\\
    &\hspace{0.8in} -J_{\vk\vp,1}\Big( -(\zeta_\text{L}-\bp) - (\zeta_\text{R}-\bk) + (\zeta_\text{R}^\dag - B_{\vk,+}) + (\zeta_\text{L}^\dag -B_{\vp,+})  \Big)\Bigg],\notag\\
    \widetilde{\Phi}_z(\i\wnu) &= \sum_{\vk\vp} \frac{|T_{\vk\vp}|^2}{4\gamma_\vk\gamma_\vp}\Bigg[ J_{\vk\vp,0}\Big(\varepsilon_{\vk\downarrow}(\varepsilon_{\vp\downarrow}+\i\wnu) -  \varepsilon_{\vk\uparrow}(\varepsilon_{\vp\uparrow}+\i\wnu) + (\zeta_\text{R}-\bk)(\zeta_\text{L}^\dag-B_{\vp,+})-(\zeta_\text{R}^\dag-B_{\vk,+})(\zeta_\text{L}-\bp)   \Big)  \notag\\
    &\hspace{0.8in} +J_{\vk\vp,1}\Big(
    \varepsilon_{\vk\uparrow}-\varepsilon_{\vk\downarrow} +
    \varepsilon_{\vp\uparrow}-\varepsilon_{\vp\downarrow} \Big) \Bigg].
  \end{align}
\end{widetext}

\end{document}